\documentclass{emulateapj}
\slugcomment{{\sc Accepted to ApJ:} September 10, 2008} 
\def\arcsec{$^{\prime\prime}$}
\newcommand\surfb{$\mathrm{mag}/\square$\arcsec}
\newcommand\kms{\rm{~km~s}$^{-1}$}

\newcommand\ha{$\rm{H}\alpha$}
\newcommand\hb{$\rm{H}\beta$}

\def\mgb{Mg~{\emph b}}

\shorttitle{Lick Indicies of Thick Disks}
\shortauthors{Yoachim \& Dalcanton}

\begin{document}

\title{Lick Indices in the Thin and Thick Disks of Edge-On Disk Galaxies}

\author{Peter Yoachim\altaffilmark{1,2} 
  \&  Julianne J. Dalcanton\altaffilmark{2,3}}

\altaffiltext{1}{Department of Astronomy and McDonald Observatory, University of Texas, Austin, TX 78712; {yoachim@astro.as.utexas.edu}}
\altaffiltext{2}{Department of Astronomy, University of Washington, Box 351580,
Seattle WA, 98195}
\altaffiltext{3}{Tom and Margo Wyckoff Fellow}

\begin{abstract}
We have measured Lick index equivalent widths to derive luminosity
weighted stellar ages and metallicities for thin and thick disk
dominated regions of 9 edge-on disk galaxies with the ARC 3.5 meter
telescope at Apache Point Observatory.  In all cases, the thick disks
are confirmed to be old stellar populations, with typical ages between
4 and 10 Gyr.  The thin disks are uniformly younger than the thick
disks, and show strong radial age gradients, with the outer regions of
the disks being younger than 1 Gyr.  We do not detect any significant
metallicity differences or $\alpha$-element enhancement in the thick
disk stars compared to the thin disk, due to the insensitivity of the
Lick indices to these differences at low metallicity.  We compare
these results to thick disks measured in other systems and to
predictions from thick disk formation models.
\end{abstract}


\section{Introduction}

Old stellar populations preserve a fossil record of a galaxy's early
formation and evolution.  In the Milky Way, the stellar thick disk and
halo have been recognized as the oldest stellar components and have
been studied extensively.  Stars in the MW thick disk are old
($\sim$8-12 Gyr), and are metal poor when compared to local thin disk
stars \citep{Reid93, Chiba00}.  Their chemical composition shows that
they are enhanced in $\alpha$-elements compared to thin disk stars,
suggesting a rapid formation timescale ($<$1 Gyr)
\citep{Reddy06,Brewer06,Bensby03,Bensby05, Prochaska00}.

Because stellar thick disks and halos are intrinsically low surface
brightness features, they have been observed in detail in only a
handful of galaxies.  Thick disks have been photometrically detected
as an excess flux at large galactic latitudes across a range of Hubble
types \citep{Dalcanton02,Burstein79, Tsikoudi79,deGrijs96, deGrijs97b,
Pohlen04,Kruit84, Shaw89, vanDokkum94, Morrison97, Wu02,Abe99,
Neeser02}, as well as detected in star counts using resolved stellar
populations from HST \citep{Seth05b,Mould05,Tikhonov05,Tikhonovo5b,Tik08}.

Three general classes of formation mechanisms have been put forward to
explain thick disks. In the first, a previously thin disk is
kinematically heated.  In this scenario, stars form in a thin disk and
increase their velocity dispersion with time.  This vertical heating
can be rapid, due to interactions and mergers
\citep{Quinn93,Walker96,Velazquez99,Chen01,Robin96,Villalobos08} or
gradual due to scattering off giant molecular clouds, spiral arms,
and/or dark matter substructure
\citep{Villumsen85,Hann02,Benson04,Hayashi06,kaz07}.  In the second
formation scenario, stars ``form thick'' with star formation occurring
above the midplane of the galaxy \citep{Brook04} or form with large
initial velocity dispersions in massive stellar clusters
\citep{Kroupa02}.  In the final class of models, thick disk stars are
directly accreted from satellite galaxies.  Numerical simulations have
shown that stars in disrupted satellite galaxies can be deposited onto
thick disk like orbits \citep{Abadi203, Martin04, Bekki01, Gilmore02,
Navarro04, Statler88,Read08}.  While these models were originally
developed to explain the origin of the MW thick disk, they should work
equally well for thick disks in other massive galaxies.

On the other hand, some of these mechanisms are likely to be less
effective in lower mass galaxies which have lower density disks,
little or no spiral structure, and fewer satellites hosting stars.
Lower mass galaxies also have different formation times, environments,
and gravitational potentials that may also lead to variation in the
mass, age, and metallicity of thick disks of low mass galaxies.

To test these formation models, detailed comparisons of thin and thick
disk properties are required across a range of galaxy masses.  In
particular, the relative ages and chemical enrichment patterns of the
thin and thick disks should differ among these formation models.  If
the thick disk is just a gradually kinematically heated thin disk,
there should be a smooth age and enrichment gradient between the two.
In contrast, if the thick disk is formed from accreted stars we should
expect the ages and metallicities of the thin and thick disks to be
only weakly correlated.  We may also expect to see variations with the
mass of the galaxies, with less massive galaxies being more
susceptible to external heating and more massive galaxies being better
able to tidally disrupt satellites.

Measuring the ages and metallicities of thick disks outside the MW has
proved to be challenging.  Ages and metallicities can be derived from
isochrone fits to thin and thick disk stars resolved with HST,
provided that the host galaxies are sufficiently close and oriented
edge-on to the line of sight.  The systems studied this way show older
populations at large scale heights but little vertical metallicity
gradient, at least for the low mass galaxies ($\rm{V}_{\rm{c}}<100$
\kms) which dominate these samples \citep{Seth05b,Tikhonov05,Mould05}.

For the systems that are farther away, only broadband colors have been
used to estimate the ages and metallicities of the thick disks.  When
thick disks are photometrically detected, they typically have very red
colors ($B-R\sim1.3-1.5$) \citep{Yoachim06, Dalcanton02, vanDokkum94}
suggestive of an old population.  However, stellar parameters are
notoriously difficult to derive from broadband colors due to the
age-metallicity degeneracy in the optical colors, and a lack of
IR-colors at the low surface brightnesses of the thick disk region.

Some progress has been made in measuring the
metallicity of the thin disk using high signal-to-noise emission lines
\citep[e.g.][]{Tremonti04, Zaritsky94, vanzee98}, and in measuring
metallicity gradients \citep[e.g.,][]{Zaritsky94,vanzee98}.  While the
emission lines studies can constrain the total chemical enrichment a
galaxy has experienced, they tell us nothing about the underlying
stellar populations in the galaxies.  Emission lines are also unable
to provide constraints on the properties of extra-planar stellar
populations, which are likely to be dominated by old, dormant stellar
populations.

To better measure the ages and metallicities of thick disks in a
larger sample of galaxies, we turn to the integrated spectrum of these
galaxies and use the Lick/IDS absorption line system to derive
luminosity-weighted stellar population properties.  The Lick indices
were originally developed for studying older stellar populations
\citep{Burstein84, Faber85}, and have been used extensively in
analyzing elliptical galaxies and globular clusters
\citep[e.g.,][]{Trager98, Trager00b, SB07}.  \citet{Worthey94} showed
that using a combination of age sensitive (Balmer lines) and
metallicity sensitive (\mgb\ and Fe) indices, one can lift the
age-metallicity degeneracy for a stellar population.  Stellar spectral
libraries have now been used to create SSP models over a large range
of metallicity and age combinations \citep{Worthey94b,Vazdekis99},
including models with variable $\alpha$-element enhancement
\citep{Thomas03}.  The tools also now exist to calculate expected Lick
indices for composite stellar populations \citep{Bruzual03}.

Despite the development of stellar synthesis codes that can be
extended to younger stellar populations, relatively few studies have
attempted to observe Lick indices in disk galaxies.  Studies using
tunable filters have been able to detect Mg and Fe index gradients in
disks \citep{Beauchamp97,Ryder05}, but have not been combined with
measurements of age-sensitive indices.  Studies of disk systems have
tended to focus on the high surface brightness bulge components
\citep{Moorthy06,Peletier07,Prugniel01,Proctor00}, and fail to reach
very far into the disks.  In the most extensive study observing Lick
indices in disk galaxies, \citet{MacArthur06} observed Lick indices in
8 galaxy disks, including several late-type galaxies out to $\sim$1
scale length.  These observations probed to $\sim$1 scale length.
However, all of these galaxies were fairly face-on, preventing thick
and thin disk components from being separated.

In this paper, we target regions of edge-on galaxies that are
dominated by either the stellar thick disk and thin components.  We then
compare ages and metallicities derived from Lick indices both
between the two components and from galaxy-to-galaxy.

\section{Observations}

The original sample of 49 galaxies was selected from the Flat Galaxy
Catalog \citep{Karachentsev93} and imaged in $B$, $R$, and $K_s$
\citep{Dalcanton00}.  \citet{Dalcanton02} used this imaging to
demonstrate the ubiquity of thick disks around late-type galaxies.  We
have since used two-dimensional decompositions of the galaxy images to
measure structural parameters for the thick and thin disks
\citep{Yoachim06}.  We have also measured kinematic properties of the
thick and thin disks using GMOS on the Gemini telescopes
\citep{Yoachim05,Yoachim07inp2}.

For this study, we selected a subset of galaxies where the photometric
decompositions suggested we would be able to isolate thin and thick
disk regions while obtaining adequate signal.  This limits
us to observing predominantly lower-mass galaxies, as
\citet{Yoachim06} found these are the galaxies with proportionally
larger thick disks.

Observations were made using the Dual Imaging Spectrograph (DIS) on
the ARC 3.5 meter telescope at Apache Point Observatory (APO) between
June 2003 and February 2006 during a total of 34 half-nights of
observing.  Although the telescope aperture is much smaller than used
for previous studies (3.5m compared to 8m), we were able to reach the
needed signal-to-noise using a novel slit design.  Spectroscopy of
extended low surface brightness objects has traditionally been limited
by the need to collect a large number of photons while maintaining
adequate spectral resolution.  However, since the Lick indices are
defined at low spectral resolution ($\sim8$ \AA) and cover a limited
wavelength range, it is possible to capture more photons by using a
much wider slit to feed a spectrograph with a higher resolution
grating.  We therefore built a custom 5\arcsec x 4$^\prime$ slit to
allow the largest possible amount of light to reach the detector.  On
the blue side, we used a grating with 1200 lines/mm and on the red
side, 830 lines/mm.  The resulting spectral resolution was 4.9(7.1)
\AA\ FWHM as measured from calibration lamps for the blue(red) chip.
The blue chip gave wavelength coverage from 4390-5430 \AA\ sufficient
to measure the age sensitive \hb\ and metallicity sensitive Mg,
Fe5270, and Fe5335 Lick indices.  The red side covered 6140-7860~\AA,
which allowed us to measure the strength of the \ha\ emission line.
The CCDs were binned by 2 in the spectral direction to reduce read
noise.  Combining a large aperture slit with a high resolution grating
is an unconventional setup, but the increased throughput of the wide
slit allows us to observe the low surface brightness regions where the
thick disk dominates.

\begin{figure}
\epsscale{.75}
\plotone{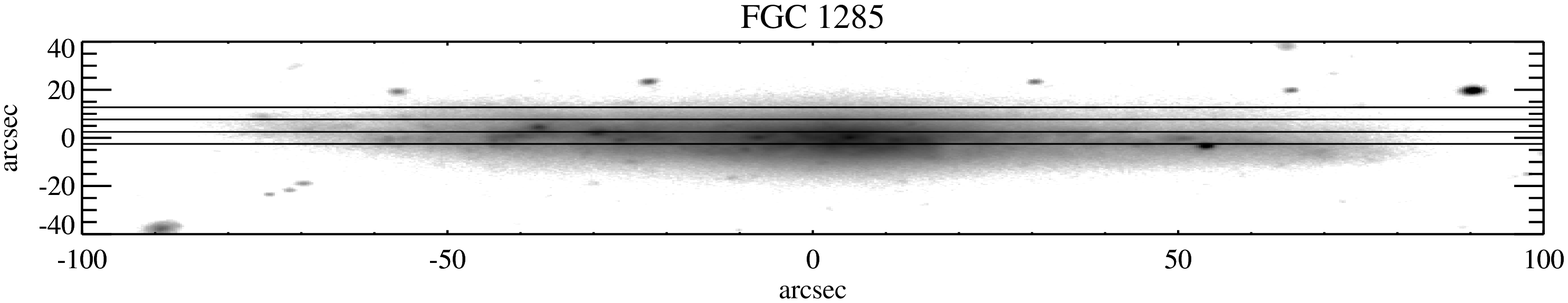}
\plotone{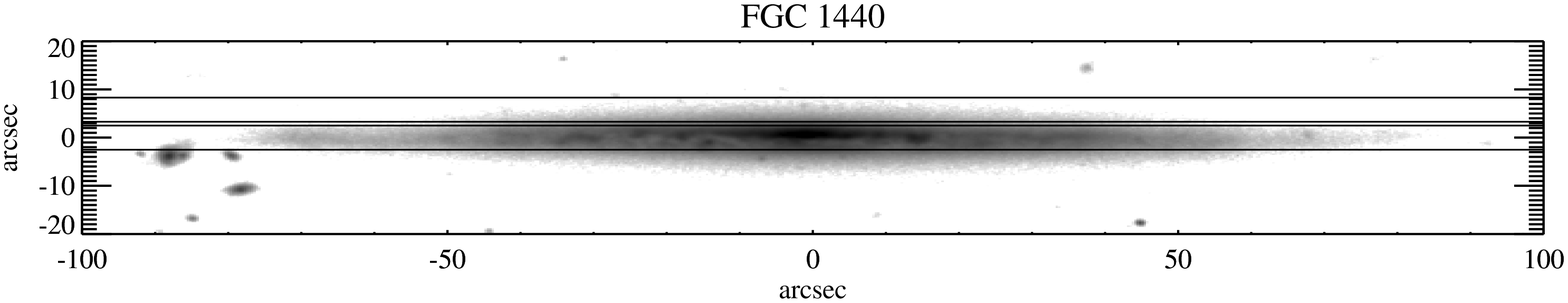}
\plotone{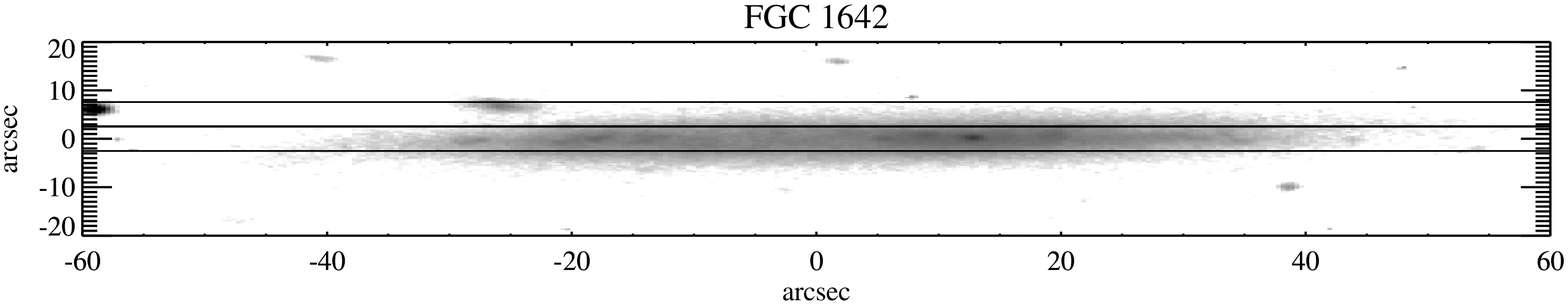}
\plotone{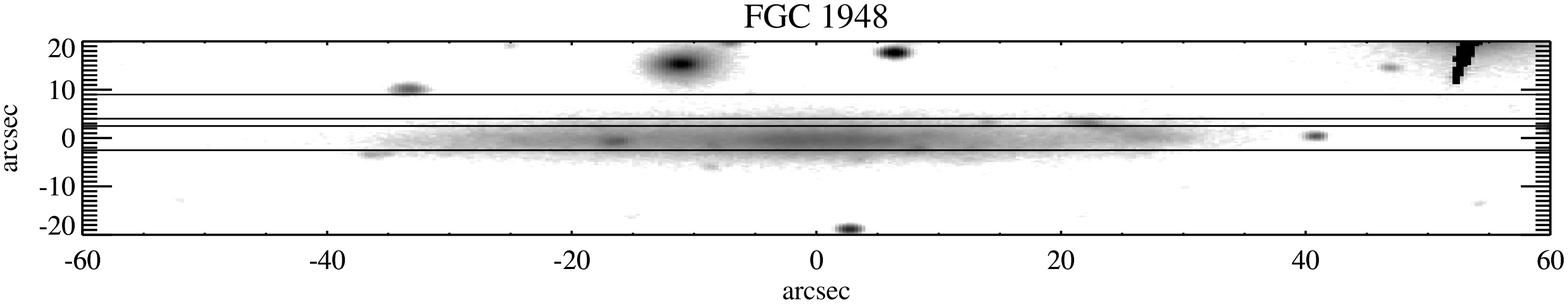}\\
\plotone{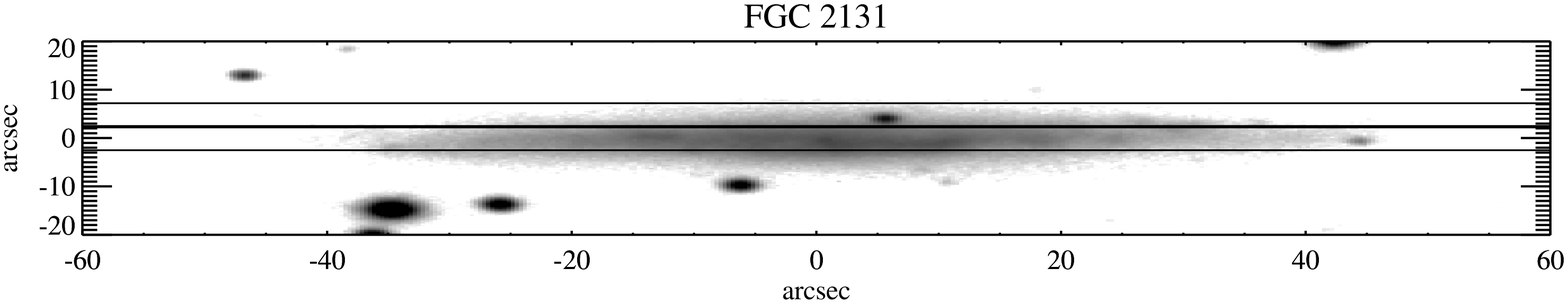}\\
\plotone{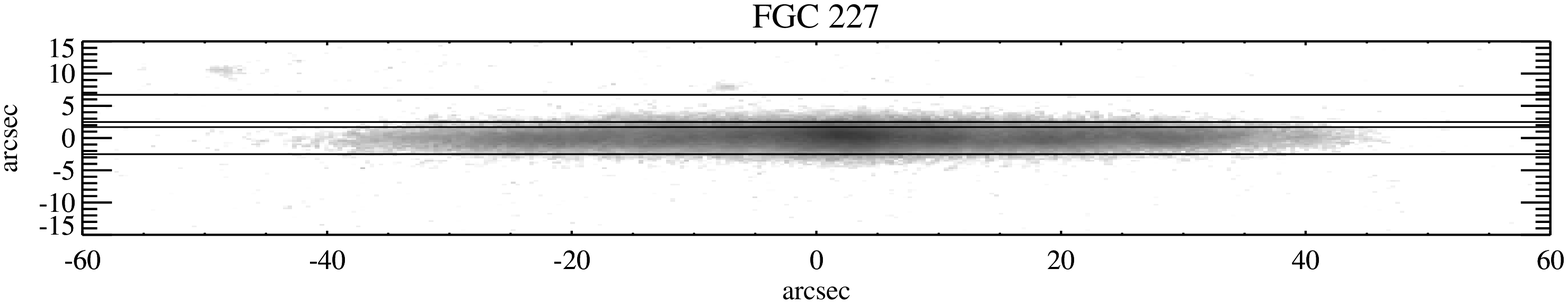}\\
\plotone{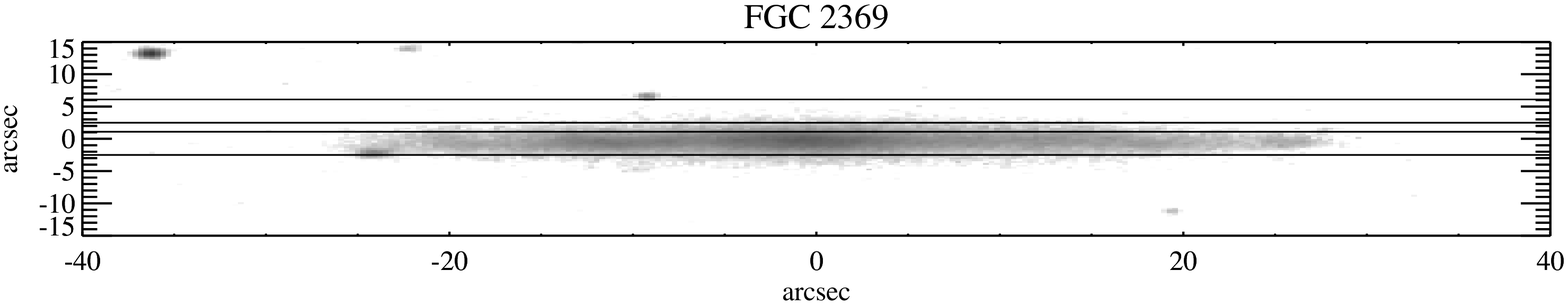}\\
\plotone{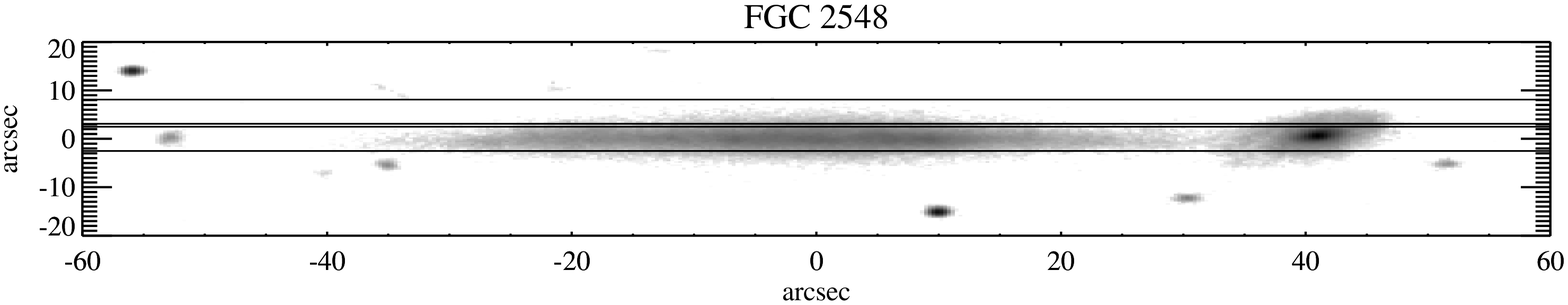}\\
\plotone{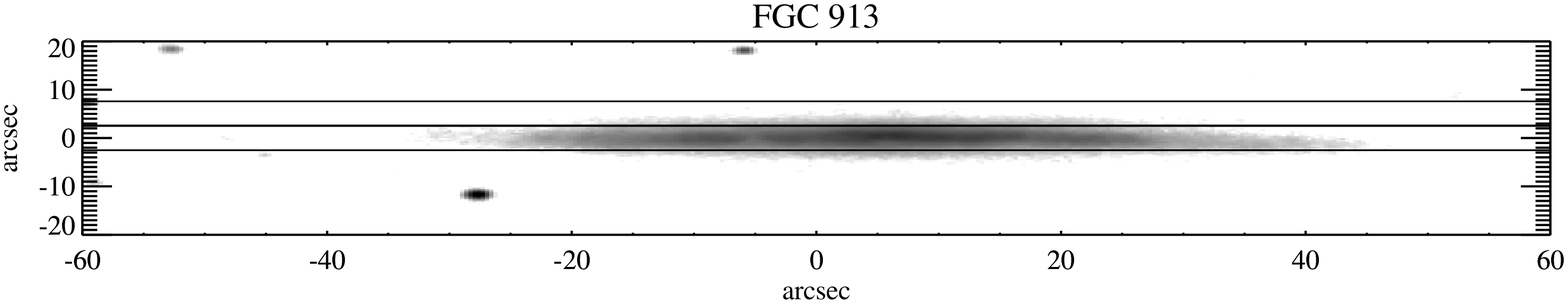}\\
\caption{$R$-band images of our galaxy sample with the APO longslit
positions overlayed.  All images are stretched to include $20 < \mu_R<
24.2$ \surfb.  \label{slit_place}}
\end{figure}

The slit camera on DIS allowed us to accurately place the slit by
centroiding nearby bright stars.  The slit camera has a plate-scale of
0.298 \arcsec/pixel, and we were typically able to place the slit
within 1 pixel of our target region.  Slit camera images were taken
throughout long exposures to ensure accurate tracking.  For
each galaxy, we gathered spectra at the midplane as well as at several
vertical scale heights above the midplane where photometric
decompositions imply the majority of flux should be supplied by the
thick disk component.  Images of the slit placement are shown in
Figure~\ref{slit_place} and the observation log is listed in
Table~\ref{fgc_obs}.

We found that there were several prominent skylines near the relevant
indices that would slowly fade for several hours after sunset.  Most
problematic was a skyline of OH at 5200 \AA, which often contaminated
the \mgb\ index.  To avoid the skylines, we took advantage of the APO
3.5m scheduling system which allocates observations in half-night
intervals and scheduled most of our observations for the second half
of the night.

\begin{deluxetable*}{ l l c c c c c c c }
\tablewidth{0pt}
\tablecaption{Edge-On Galaxy Observations \label{fgc_obs}}
\tablehead{  
\colhead{Galaxy } & \colhead{Date Range} & \multicolumn{2}{c}{Midplane} & \multicolumn{2}{c}{Offplane} & \multicolumn{2}{c}{Offsets} & \colhead{Scale Length}\\
\colhead{ FGC}  & \colhead{Observed} &\colhead{exposures} & 
  \colhead{total time (min)} & \colhead{exposures} & \colhead{total time (min)}  & \colhead{arcsec} & \colhead{kpc\tablenotemark{1}} & \colhead{arcsec}}
\startdata
227  & 10/2004 to 10/2005  &    7 &    150 &     20 &    480  &4.2  & 1.7 & 10.2 \\
913  & 02/2004 to 03/2004  &    7 &     80 &     32 &    698 &5.1  &  1.5 & 9.0 \\
1285 & 02/2004 to 04/2004  &    8 &    105 &     18 &    510 & 10.2 & 0.9 & 19.7 \\
1440 & 03/2004 to 02/2006  &    4 &     80 &     10 &    270 & 5.8 & 2.0  & 15.9 \\
1642 & 06/2003 to 05/2005  &    5 &    100 &      5 &    150 & 5.1 &  0.9 & 12.5 \\
1948 & 06/2003 to 05/2005  &    5 &     95 &     15 &    405& 6.5 &  1.2  & 12.3 \\
2131 & 06/2005 to 06/2005  &    3 &     45 &     17 &    338&  4.7&  1.0  & 10.0 \\
2369 & 10/2004 to 08/2005  &    5 &    105 &     25 &    521 & 3.6 &  1.0 & 8.7 \\
2548 & 10/2003 to 10/2003  &    5 &     75 &     27 &    536 & 5.6 &  1.5 & 9.9  \\
\enddata
\tablenotetext{1}{Distances from \citet{Kara00}}
\end{deluxetable*}

\begin{deluxetable}{ l c c}
\tablewidth{0pt}
\tablecaption{Elliptical Galaxies Observed \label{ell_obs}}
\tablehead{\colhead{Galaxy} & \colhead{exposures} &\colhead{ total time (min)} }
\startdata
NGC 1453 & 2 &   25.  \\
NGC 1600 & 2 &   20.  \\
NGC 2778 & 3 &    9.  \\
NGC 3379 & 1 &    5.  \\
NGC 5638 & 2 &   10.  \\
NGC 6127 & 2 &   20.  \\
NGC 6702 & 2 &   10.  \\
NGC 6703 & 2 &   20.  \\
NGC 7052 & 1 &    5.  
\enddata
\end{deluxetable}


\section{Data Reduction}
Data were processed with standard IRAF/PyRAF routines along with
several custom scripts written in IDL.  Reduction steps for the red
and blue side were identical unless otherwise noted.  All the frames
were bias corrected by subtracting the mean from the overscan region.
Any residual bias structure was removed by subtracting a bias frame
constructed from 5-10 bias exposures taken every night.  For one half
night (Feb 11, 2004 observations of FGC 913), the blue chip suffered
from 60 Hz noise which resulted in diagonal streaks across the images.
This pattern was removed by shifting and subtracting a single bias
frame that had similar variation in the readout pattern.  The images
were flat-fielded using spectra of a quartz lamp.  Spectra of the
twilight sky showed no need for an illumination correction.  The
spectra were wavelength calibrated using He, Ne, and Ar arc lamps
along with night-sky emission lines \citep{Oster96}.  Arc spectra were
taken interspersed with the science exposures throughout the night,
approximately every hour.  Stellar spectra were used to correct the
spatial distortions (tilt) in the observations.  Observations of
standard stars and standard atmospheric extinction curves were used to
flux calibrate the spectra.  Few of the observations were made in
photometric conditions, thus our flux calibration is primarily used as
a first order removal of the instrumental sensitivity profile.
Because we are primarily interested in measuring equivalent widths,
the exact flux normalization is not crucial.  The sky was subtracted
using a second order polynomial fit to regions dominated by the sky.
The spectra were finally corrected for motion relative to the Local
Standard of Rest, scaled to a common flux level, spatially aligned,
and combined rejecting cosmic ray hits.

Our final spectra have a spatial scale of 0.42 arcsec/pixel on both
the red and blue chips, and a wavelength solution of 1.24 \AA/pixel in
the blue and 1.68 \AA/pixel in the red.  Measurement of arc lamp lines
showed a FHWM resolution of $4.9$ \AA\ on the blue chip and 7.1 \AA\
on the red chip.  \citet{Worthey97} report that the resolution (FWHM)
of the original Lick indices of interest as 8.4 \AA.  We therefore
broaden our blue spectra with a Gaussian kernel with $\sigma=2.9$ \AA\
to match the Lick system resolution.

Systematic and rotational velocities were removed by cross-correlating
a (logarithmically-binned) stellar template plus Gaussian emission
lines.  The shifts were accurate to within 1 pixel ($\sim 1.2$ \AA).
The midplane rotation curve was used for both the midplane and
offplane spectra, as any difference between the two should be small at
our resolution \citep{Yoachim07inp2}.

After the 2D spectra were broadened to match the Lick resolution, any
foreground stars were masked and 1D spectra were extracted by summing
in the spatial direction.  For both the midplane and offplane, we
extracted spectra from the region of $\pm$1 scale length ($h_R$), as
measured in \citet{Yoachim05}.

We tested how the SNR of our spectra affects the accuracy of the
measured Lick indices.  Adding artificial noise to template stellar
spectra, we find the RMS error in a measured Lick index scales with
S/N as $\sigma_{index}\sim 9(SNR)^{-1}$\AA.  Therefore, to achieve an
Lick EW uncertainty of $\pm0.2$\AA\ requires a SNR of $\sim45$ per
\AA, which agrees with previous determinations by
\citet{Trager00}.

Throughout our analysis, we will calculate the uncertainties in our
derived ages and metallicities using the SNR of the extracted spectra.
As an additional check, we have extracted 1D spectra from $0<R/h_R<1$
and $-1<R/h-R<0$ to find the variance in the Lick indices from one side
of the galaxy to the other and note cases where there are large
discrepancies.  This procedure allows us to flag systems where
systematic errors are likely to dominate over random
uncertainties.


\begin{figure}
\plotone{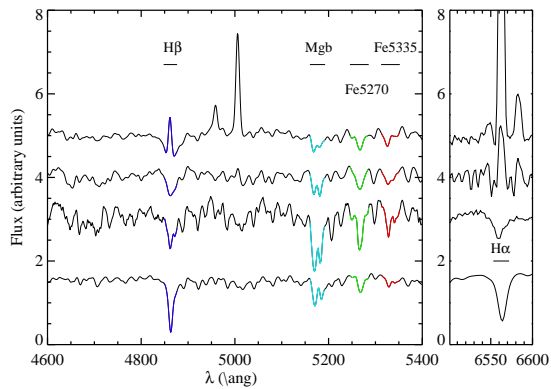}
\caption{Examples of our blue and red spectra.  From top to bottom:
Midplane of FGC 1285, Offplane of FGC 1285, Elliptical galaxy NGC
5638, Lick standard star HD114762.  All the blue spectra have been smoothed
to a resolution of 8.4 \AA\ and the Lick indices of interest have been
labeled.
\label{example_spec}}
\end{figure}


\section{Moving Onto the Lick System}

\begin{figure}
\plotone{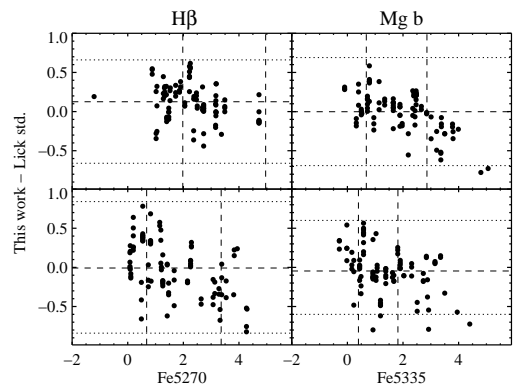}
\caption{Comparison of Lick indices measured with DIS compared to the
published equivalent widths in \citet{Worthey94}.  Dashed horizontal
lines show the mean offsets while the dotted horizontal lines show the
average Worthey et al. $\pm 3 \sigma$ uncertainties.  Dashed vertical
lines show the range of the line indices measured in our galaxy sample.
Corresponding means and scatters are given in Table~\ref{zp_table}.
\label{resids2}}
\end{figure}

While we have matched the Lick resolution reported in
\citet{Worthey97}, this is not enough to ensure that we are on the
Lick system.  Because the Lick indices were originally defined from
spectra that had not been flux-calibrated, additional corrections are
needed.  To do so, we made 144 observations of 37 unique Lick standard
stars.  Stars for which we have repeat observations have a mean RMS
error of 0.09 \AA\ for each index.  Our standard star EWs are compared
to values listed in \citet{Worthey94}, and are shown in
Figure~\ref{resids2}.  The derived zero point corrections are
listed in Table~\ref{zp_table}.  In general, the agreement is quite
good, with a typical Gaussian scatter of 0.26 \AA\ and little systematic
offset.  The notable exception is for \mgb, for which the APO system
measures low EW for metal rich systems with \mgb$>3$ \AA.  We do not
explicitly correct for this offset as we are observing metal poor
galaxies which have \mgb\ indices $<2$ \AA.  We also compare our standard
star EWs with stars in common with \citet{Schiavon06} and find
a spread of 0.2-0.4 \AA.  \citet{Schiavon06} points out that
measurements of EWs of bright stars are disturbingly inconsistent, and
that there can be surprisingly large variations between observations.
\citet{Schiavon06} argue that these offsets are caused by errors in
the flat-fielding, which dominate the errors of the bright stars,
unlike fainter galaxy spectra which are background limited.


\begin{deluxetable}{c c c c c}
\tablewidth{0pt}
\tablecaption{Zero point conversions \label{zp_table}}
\tablehead{index& H$\beta$ & \mgb & Fe 5270 & Fe 5335 }
\startdata
 zero point\tablenotemark{1} (\AA) & 0.13 & 0.00 & -0.01 & -0.05  \\
 RMS (\AA) & 0.23 &  0.26 &  0.35 &  0.29 
\enddata
\tablenotetext{1}{Value subtracted from our measured indices to move onto the Lick system}
\end{deluxetable}


\begin{figure}
\plotone{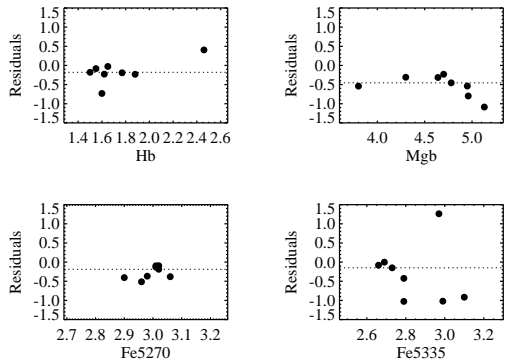}
\caption{ Lick indices measured with APO compared with values
published in \citet{Trager00}.  Dashed lines show the median offset
between our measurements and those in \citet{Trager00}.  The agreement
is good, even with our larger slit and lack of velocity dispersion
correction.  The large errors for the Fe5335 index values are caused
by the feature approaching the DIS dichroic cut-off.
\label{ell_resids}}
\end{figure}


We also observed a sample of elliptical galaxies which have reported
Lick measurements in the literature.  The ellipticals were observed
when light cloud cover made observing faint regions of disks
impossible, or during brief periods when the primary targets were not
visible.  We observed 9 galaxies in common with \citet{Trager00}
listed in Table~\ref{ell_obs}.  Our indices are not directly
comparable to the ones listed in \citet{Trager00}, as we used our
large 5\arcsec\ slit which samples more flux from the outer regions of
the galaxies, and we have not replicated the corrections for velocity
dispersion and emission line fill-in as done in \citet{Trager00}.
Thus, some amount of scatter is expected.  The residuals between our
measurements and those in \citet{Trager00} are plotted in
Figure~\ref{ell_resids}.  For the H$\beta$, \mgb, and Fe 5270 indices,
we measure RMS scatters of 0.16-0.31 \AA\ around the \citet{Trager00}
values.  There is a much larger scatter for Fe 5335 because the
feature approaches the DIS dichroic cut-off for galaxies with a
redshift greater than 4000 \kms\, and becomes very low
signal-to-noise.

\subsection{Emission Line Removal}
Unlike observations of gas-poor ellipticals and globular clusters, we must
remove any emission lines that contaminate the index passbands before
we can derive ages and metallicities from the absorption features.
The removal of emission lines is not trivial.  We would like to remove
the \hb\ emission, but to do so we would need to know the underlying
shape of the stellar continuum.  This shape in turn depends strongly
on the stellar population's age and metallicity, which is what we set
out to measure.  We therefore must turn to other parts of the spectrum
to estimate the amount of \hb\ emission.

Lick indices, particularly \hb, are often contaminated with emission
lines.  \citet{Gonzalez93} popularized using the correction
\hb$_{emission}$=0.7[O\,{\sc iii}]$\lambda$5007.  This correction was derived
by fitting a stellar template to the underlying spectrum of their
elliptical galaxies to isolate the emission feature.  Unfortunately,
the shape of the template can determine the magnitude of the emission
correction and the ratio of $F_{\rm{H}\beta}/F_{[\rm{O}III]}$ is
highly metallicity sensitive \citep[e.g.,][]{Kewley02}, making this correction
inappropriate for low metallicity systems with varying stellar ages.
\citet{Moorthy06} make a correction by fitting a Gaussian to the
emission peak inside the \hb\ Lick index.  This correction probably
underestimates the \hb\ correction since the Gaussian fit will not be
sensitive to the fraction of \hb\ emission that has filled the
absorption feature.  This correction is also very sensitive to the
resolution of the measured spectrum.  When we try to fit a Gaussian to
the emission feature after broadening, we find that the correction is
0.4 \AA\ smaller than when we make the correction before broadening.
\citet{MacArthur06} takes the most extreme measure of masking regions
where the spectrum is contaminated by emission lines.

Using the [O\,{\sc iii}] emission line to estimate \hb\ is a fine
approximation when the \hb\ emission correction is small, as in
elliptical galaxy spectra.  However, our midplane spectra are clearly
dominated by the \hb\ emission (Figure~\ref{example_spec}), and we
must use a more accurate technique to remove the emission.  To do so,
we measure the EW of the \ha\ emission line from the spectra taken
simultaneously with the red DIS spectrograph and use this to estimate
the EW of the \hb\ emission line.

Our procedure for removing the \hb\ emission is as follows.  We first
assume case B recombination with
$F_{\rm{H}\alpha}=2.86F_{\rm{H}\beta}$ \citep{Osterbrock89}.  This
correction is used in \citet{Rampazzo05} and \citet{Denicol05} when
measuring Lick indices in elliptical galaxies.  However, because the
\ha\ and \hb\ emission lines were measured on different CCDs, we are
hesitant to use the measured flux ratios.  While both CCDs are
calibrated using the same flux standard stars, the subsequent scaling
and co-adding of frames was done independently and could skew the
absolute flux calibration between the two.  Instead, we make
corrections based on equivalent widths as follows. 

The definition of the EW is
\begin{equation}
{\rm{EW}}=\int_{\lambda_1}^{\lambda_2}\left(1-\frac{F_{I\lambda}}
{F_{C\lambda}}\right)d\lambda 
\end{equation}
where $F_{I\lambda}$ is the spectrum in the index passband and $F_{C\lambda}$ is the continuum spectrum calculated from the flanking pseudo-continuum regions.  Our goal is to remove the contaminating emission feature to recover the equivalent-width of the underlying Lick absorption feature.
\begin{equation}
\rm{EW_{observed}}=\rm{EW_{Lick}}+\rm{EW_{emission}}
\end{equation}

    In the case where the continuum is constant, the EW from the emission alone is given by
\begin{equation}
{\rm{EW}_{emission}}=\bigtriangleup\lambda-\frac{F_{em}+
F_{C\lambda}\bigtriangleup\lambda}{F_{C\lambda}}=
-\frac{F_{em}}{F_{C\lambda}}
\end{equation}
where $F_{em}$ is the integrated flux in the emission line and $F_{C\lambda}$ is the continuum level.  We can then write the expected EW ratio for the Balmer emission lines as
\begin{equation}
\frac{\rm{EW(H}\alpha)_{em}}{\rm{EW(H}\beta)_{em}}=
\frac{F_{\rm{H}\alpha}}{F_{\rm{H}\beta}}
\frac{F_{C\lambda,\rm{H}\beta}}{F_{C\lambda,\rm{H}\alpha}}.
\end{equation}
This in turn can be modified to account for differential extinction between the stars and HII regions.

\begin{equation}
\frac{\rm{EW(H}\alpha)_{em}}{\rm{EW(H}\beta)_{em}}=
\frac{F_\alpha}{F_\beta}
\frac{F_{C\lambda,\beta}}{F_{C\lambda,\alpha}} 
\frac{10^{0.4E(B-V)_{gas}k(\rm{H}\beta) - k(\rm{H}\alpha)}}{10^{0.4E(B-V)_{stars}k(\rm{H}\beta) - k(\rm{H}\alpha)}}
\end{equation}
\citet{Calzetti01} lists $ k ({\rm{H}}\beta) - k ({\rm{H}}\alpha)=1.163$ and finds that  $E(B-V)_{stars}=0.44E(B-V)_{gas}$.  The difference in reddening is due to the geometrically clumpy distribution of HII regions compared to the smoother distribution of stars.  Finally, we get the relation
\begin{equation}
\frac{\rm{EW(H}\alpha)_{em}}{\rm{EW(H}\beta)_{em}}=
\frac{F_{\rm{H}\alpha}}{F_{\rm{H}\beta}}
\frac{F_{C\lambda,\rm{H}\beta}}{F_{C\lambda,\rm{H}\alpha}} 
10^{0.26E(B-V)_{gas}}.
\end{equation}

For the offplane spectra, we assume there is negligible dust
extinction, a fairly flat continuum level
($F_{C\lambda,\rm{H}\beta}\approx F_{C\lambda,\rm{H}\alpha}$), and
case B recombination, resulting in the standard correction of
${\rm{EW(H}}\beta)_{em}=y {\rm{EW(H}}\alpha)_{em}$ with $y=2.86$.  In
the case of the midplane spectra, there is a younger bluer stellar
population and the possibility of dust extinction.  We adopt case B
recombination, a continuum ratio of 1.1, and $E(B-V)\sim0.1$ resulting
in $y=3.3$.  Our value of $E(B-V)$ is lower than typical values for
the centers of edge-on disks (\citet{Matthews99} find $E(B-V)\sim0.2$
in an edge-on disk similar to our sample galaxies), however, light off
the midplane and at larger radii experience significantly less
extinction.  For elliptical galaxies identified as dusty,
\citet{Denicol05} adopt a correction of $y=3.0$.

These adopted corrections give results consistent with what we expect
from the broadband colors and SSP models.  Most of the corrected \hb\
Lick EWs fall within the range expected from the model grids, and we
find younger ages where the integrated light is blue and older
ages where it is red.  In \S\ref{elcorr}, we calculate how a 10\%
change in our adopted value of $y$ would propagate to the derived ages
and metallicities.  This uncertainty corresponds to extinction values $0.08 <
E(B-V) < 0.42$ or continuum ratios $1.05
<F_{C\lambda,\rm{H}\alpha}/F_{C\lambda,\rm{H}\beta} < 1.3$.

While we consider how dust affects the emission line EW ratios, the
Lick absorption features themselves are fairly insensitive to dust
\citep{MacArthur05} and require no extra corrections even if the
galaxies are dusty.


\begin{deluxetable*}{ r c c c c c c c   }
\tabletypesize{\small}
\tablewidth{0pt}
\tablecaption{Measured Lick indices in the range $-1 < h_R <1$\label{lick_table}}
\tablehead{  
\colhead{Galaxy } & \colhead{SNR/Pixel} &  \colhead{\hb$_{\rm{raw}}$} & \colhead{\ha} & \colhead{\hb$_{\rm{corr}}$} & \colhead{\mgb} &\colhead{Fe 5270} & \colhead{Fe 5335}
}
\startdata
FGC1285 midplane &  272.18 &   -0.02$\pm 0.0$ &   -9.82$\pm 0.1$ &    2.92$\pm 0.0$ &    1.16$\pm 0.0$ &
   1.45$\pm 0.0$ &    1.54$\pm 0.0$  \\
 offplane &   91.21 &    1.68$\pm 0.1$ &   -1.43$\pm 0.1$ &    2.18$\pm 0.1$ &    1.89$\pm 0.1$ &    1.21$\pm 0.1$ &
   1.27$\pm 0.1$  \\
FGC1440 midplane &  200.25 &   -0.92$\pm 0.0$ &  -11.83$\pm 0.1$ &    2.62$\pm 0.0$ &    2.29$\pm 0.0$ &
   1.83$\pm 0.1$ &    1.23$\pm 0.1$  \\
 offplane &   36.24 &    0.59$\pm 0.2$ &   -4.28$\pm 0.1$ &    2.09$\pm 0.2$ &    2.63$\pm 0.2$ &    3.36$\pm 0.3$ &
   6.75$\pm 0.5$  \\
FGC1642 midplane &   76.15 &    0.35$\pm 0.1$ &  -11.28$\pm 0.1$ &    3.73$\pm 0.1$ &    0.78$\pm 0.1$ &
   0.68$\pm 0.1$ &    0.40$\pm 0.2$  \\
 offplane &   60.95 &   -0.24$\pm 0.1$ &   -8.29$\pm 0.1$ &    2.66$\pm 0.1$ &    1.45$\pm 0.1$ &    1.45$\pm 0.2$ &
   0.84$\pm 0.3$  \\
FGC1948 midplane &   77.36 &   -2.30$\pm 0.1$ &  -22.40$\pm 0.1$ &    4.41$\pm 0.1$ &    0.53$\pm 0.1$ &
  -0.16$\pm 0.1$ &    0.87$\pm 0.2$  \\
 offplane &   15.83 &    0.05$\pm 0.5$ &   -8.99$\pm 0.1$ &    3.19$\pm 0.5$ &   -0.20$\pm 0.6$ &   -0.31$\pm 0.6$ &
  -1.35$\pm 0.6$  \\
FGC2131 midplane &   98.92 &   -3.82$\pm 0.1$ &  -29.31$\pm 0.1$ &    4.96$\pm 0.1$ &    0.68$\pm 0.1$ &
   0.94$\pm 0.1$ &    1.82$\pm 0.1$  \\
 offplane &   14.71 &    0.16$\pm 0.5$ &   -8.13$\pm 0.1$ &    3.01$\pm 0.5$ &    1.47$\pm 0.6$ &    1.51$\pm 0.7$ &
   0.88$\pm 0.7$  \\
FGC227 midplane &  104.72 &   -0.58$\pm 0.1$ &  -16.33$\pm 0.1$ &    4.30$\pm 0.1$ &    0.77$\pm 0.1$ &
   1.24$\pm 0.1$ &    0.87$\pm 0.1$  \\
 offplane &   21.54 &   -1.60$\pm 0.3$ &  -10.25$\pm 0.1$ &    1.98$\pm 0.3$ &    2.03$\pm 0.4$ &    2.45$\pm 0.4$ &
   8.97$\pm 0.6$  \\
FGC2369 midplane &   51.41 &   -0.70$\pm 0.1$ &  -14.38$\pm 0.1$ &    3.60$\pm 0.1$ &    1.05$\pm 0.1$ &
   0.99$\pm 0.2$ &   -0.31$\pm 0.3$  \\
 offplane &    9.37 &    1.64$\pm 0.9$ &   -4.84$\pm 0.1$ &    3.33$\pm 0.9$ &    3.43$\pm 0.9$ &    0.19$\pm 0.9$ &
   0.33$\pm 1.1$  \\
FGC2548 midplane &   81.29 &    0.64$\pm 0.1$ &  -11.25$\pm 0.1$ &    4.00$\pm 0.1$ &   -0.12$\pm 0.1$ &
   0.99$\pm 0.1$ &   -0.04$\pm 0.2$  \\
 offplane &   28.38 &    0.67$\pm 0.2$ &   -3.66$\pm 0.1$ &    1.95$\pm 0.2$ &    0.12$\pm 0.3$ &   -0.32$\pm 0.4$ &
  -1.74$\pm 0.6$  \\
FGC913 midplane &   71.87 &   -2.43$\pm 0.1$ &  -22.14$\pm 0.1$ &    4.20$\pm 0.1$ &    1.37$\pm 0.1$ &
   1.30$\pm 0.2$ &   -0.79$\pm 0.3$  \\
 offplane &   35.48 &   -0.57$\pm 0.2$ &   -9.45$\pm 0.1$ &    2.73$\pm 0.2$ &    2.86$\pm 0.2$ &    1.19$\pm 0.3$ &
  -1.94$\pm 0.4$  

\enddata
\tablecomments{All Lick index values are measured in angstroms.  Negative values indicate emission.}

\end{deluxetable*}


\subsection{Deriving Ages and Metallicities}

Ideally, we would fit full star formation and chemical evolution
histories for our galaxies.  However, we have chosen to focus on only
4 of the most prominent Lick indices, limiting the total number of
parameters we can hope to fit.  We therefore choose to interpret the
spectra using SSP models.  While our galaxies are clearly more
complicated than single-burst stellar populations (given that we see
old stellar populations along with signs of current star formation),
the SSP models will still give a reasonable first-order approximation
of the luminosity weighted mean ages, metallicities, and
$\alpha$-element enhancements.

To convert our measured indices to ages we start with the model grids
of \citet{Thomas03} and interpolate them to a finer grid with age
steps of $\delta t=0.1$ Gyr and metallicity steps of
$\delta$[Z/H]=0.01.  We exclude the very young age grid points (age $<
0.1$ Gyr) as the Lick indices expected for such young ages are
degenerate with older populations.  To calculate the uncertainties, we
interpolate ages and metallicities for 100 Lick index pairs
distributed according to the equivalent width uncertainties.  We also
test for possible systematic errors caused by incorrect emission line
corrections and calculate how a 10\% error in the emission line
correction for H$\beta$ propagates to errors in age and metallicity.

\begin{figure}
\epsscale{.6}
\plotone{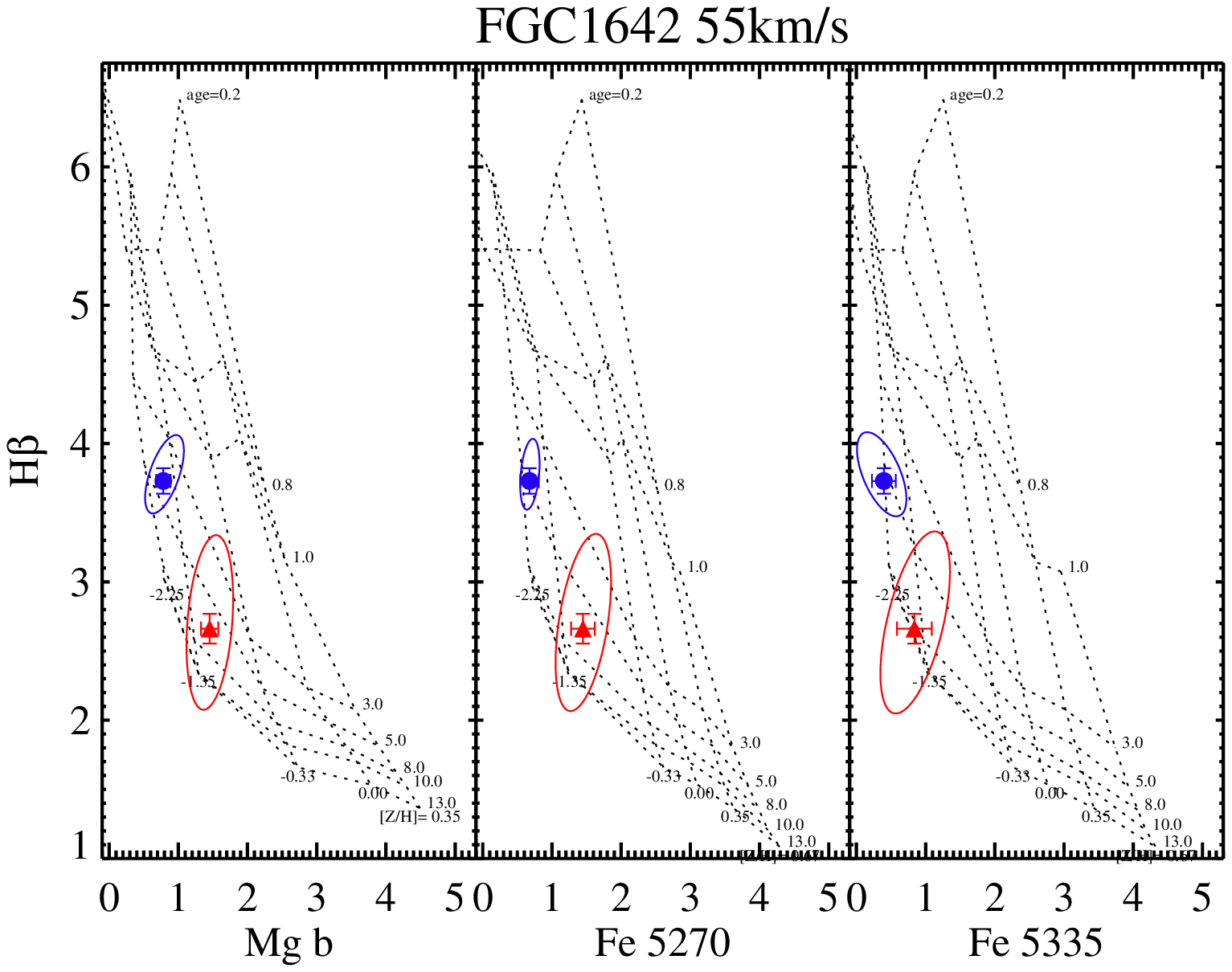}
\plotone{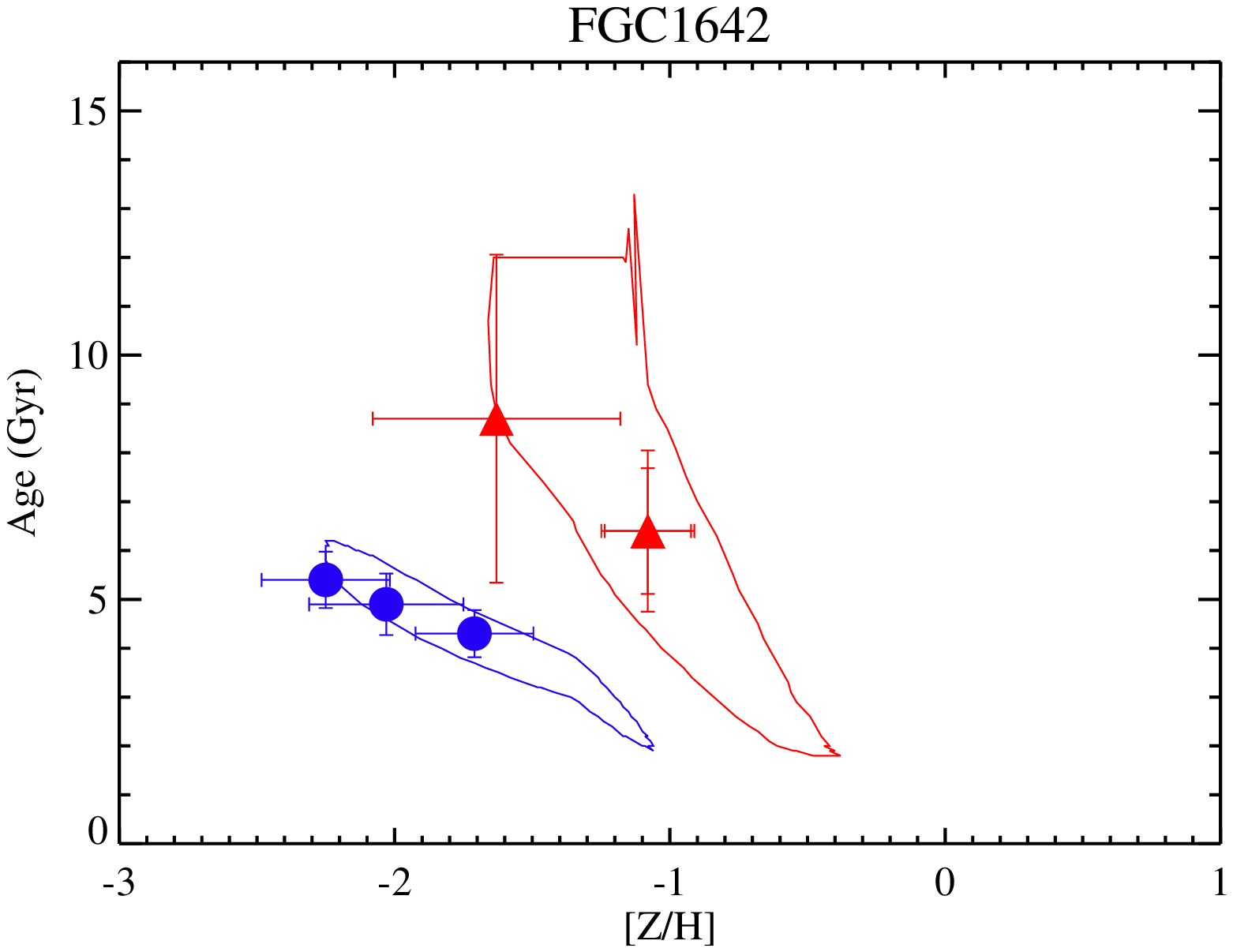}
\caption{Lick index measurements for our observations.  Top plots:
Model grids from \citet{Thomas03} along with our observed points.
Blue points are from midplane observations and red points are for
offplane measurements measured from spectra extracted from $R<|h_R|$.
Ellipses are drawn that encompass indices measured from the regions $0
< R < h_R$ and $-h_R < R < 0$ .  Small ellipses imply the measured
indices are robust, while large ellipses show that the spectra on
either side of the galaxy are not particularly consistent.  Bottom
Plots: The interpolated ages and metallicities for all the above
indices.  Large filled points are used for spectra from the region
$-h_R < R < h_R$ that fall inside the model grids.  The ellipses from
the \hb-\mgb\ plane are propagated to the bottom plots.
Throughout, blue circles are used for midplane observations while red
triangles are used for offplane observations.  Error bars on the
points show the statistical uncertainties based on the SNR of the
spectra.
  \label{grid1} }
\end{figure}

Figure~\ref{grid1} shows the Lick indices measured for each of our
galaxies, placed on the model grids of \citet{Thomas03}, assuming no
$\alpha$-enhancement.  For each galaxy, we calculate the age and
metallicity by interpolating onto grids of \hb+\mgb, \hb+Fe 5270, and
\hb+Fe 5335.  

\begin{figure}
\plotone{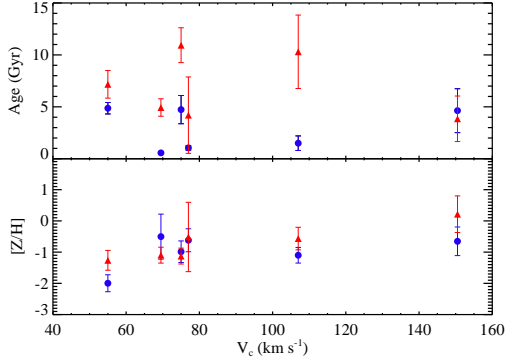}
\caption{Average ages and metallicities as measured by Lick indices \hb, \mgb, and Fe 5270.  Blue circles show midplane (thin disk) observations while red triangles show offplane (thick disk) observations. \label{age_met}}
\end{figure}

\begin{figure}
\plotone{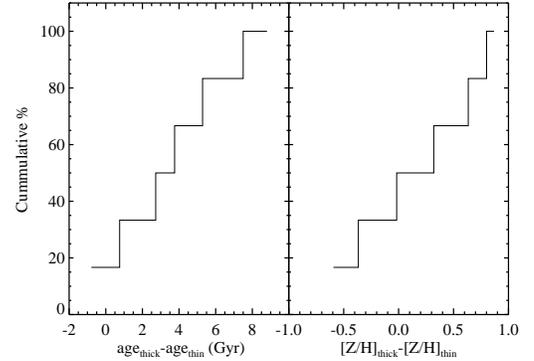}
\caption{Cumulative distribution plots showing the differences between the thin disk and thick disk ages and metallicities.  Positive ages mean the thick disk is older while negative metallicity differences mean the thick disk is metal poor in comparison to the thin disk. \label{hists}}
\end{figure}

Three of the galaxies (FGC 1948, 2369, and 2548) have Lick indices
that consistently fall far outside the model grids.  This is not
surprising, as the offplane SNR for FGC 1948 and 2369 were so low we
should not have expected to measure Lick indices
(Table~\ref{lick_table}), and FGC 2548 is only on the cusp of having
adequate SNR.  For the remaining six galaxies, we plot the average
interpolated ages and metallicities for the midplane and offplane in
Figure~\ref{age_met} and plot cumulative distributions of the age and
metallicity differences between the thin and thick disk components in
Figure~\ref{hists}.  For consistency, we use only the metallicities
measured from the \mgb\ and Fe 5270 indices because several galaxies
have very low SNR at the Fe 5335 index.

For the 6 galaxies with high SNR, we find the thick disks have a
median age of 7.2 Gyr and metallicity of [Z/H]=-0.6.  The thin disks
have a similar median metallicity as expected for their low mass, but
are uniformly much younger, with a median age of 4.6 Gyr.  This age
determination for the thin disk is derived after making very large
emission line corrections and represents a flux-weighted average of
all the stars within a radius of one scale length.  It is certainly
possible that the oldest central regions of the thin disk contain
stars whose ages are similar to those of the thick disks.  We return
to these results in \S\ref{sec_radgrad} below.


\clearpage
{
\plotone{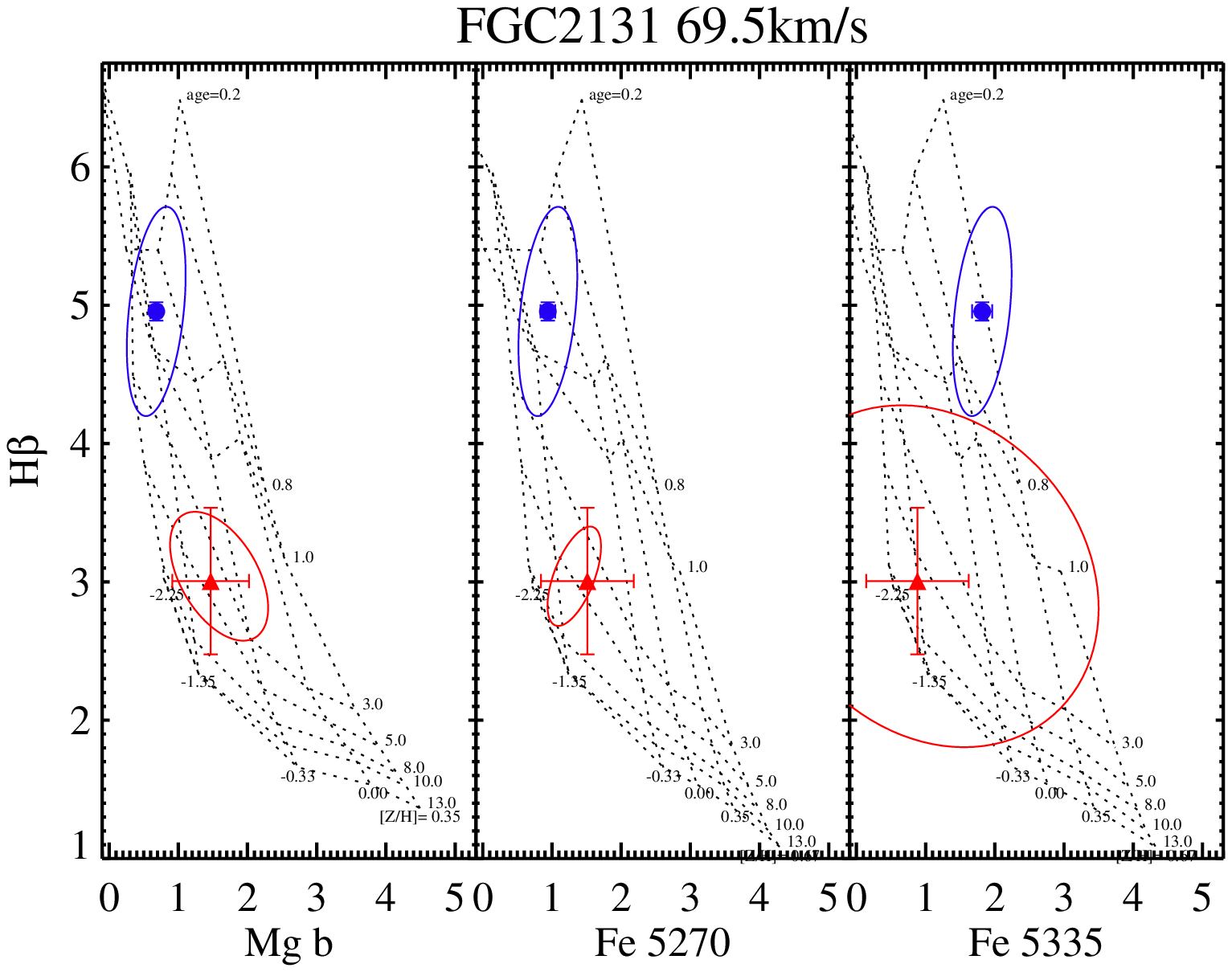}\\
\plotone{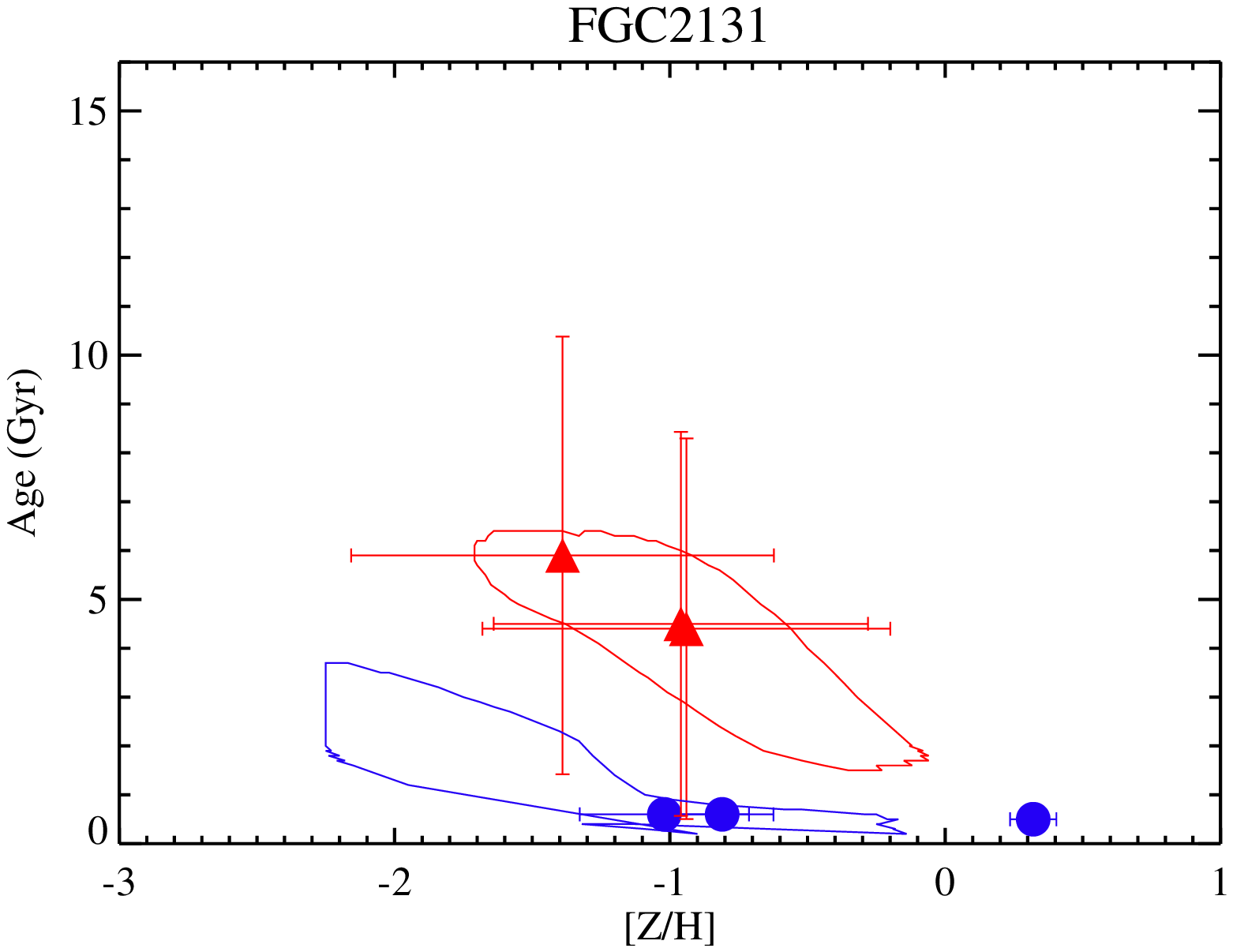}\\
}
\centerline{Fig. 5. --- Continued.}
{
\plotone{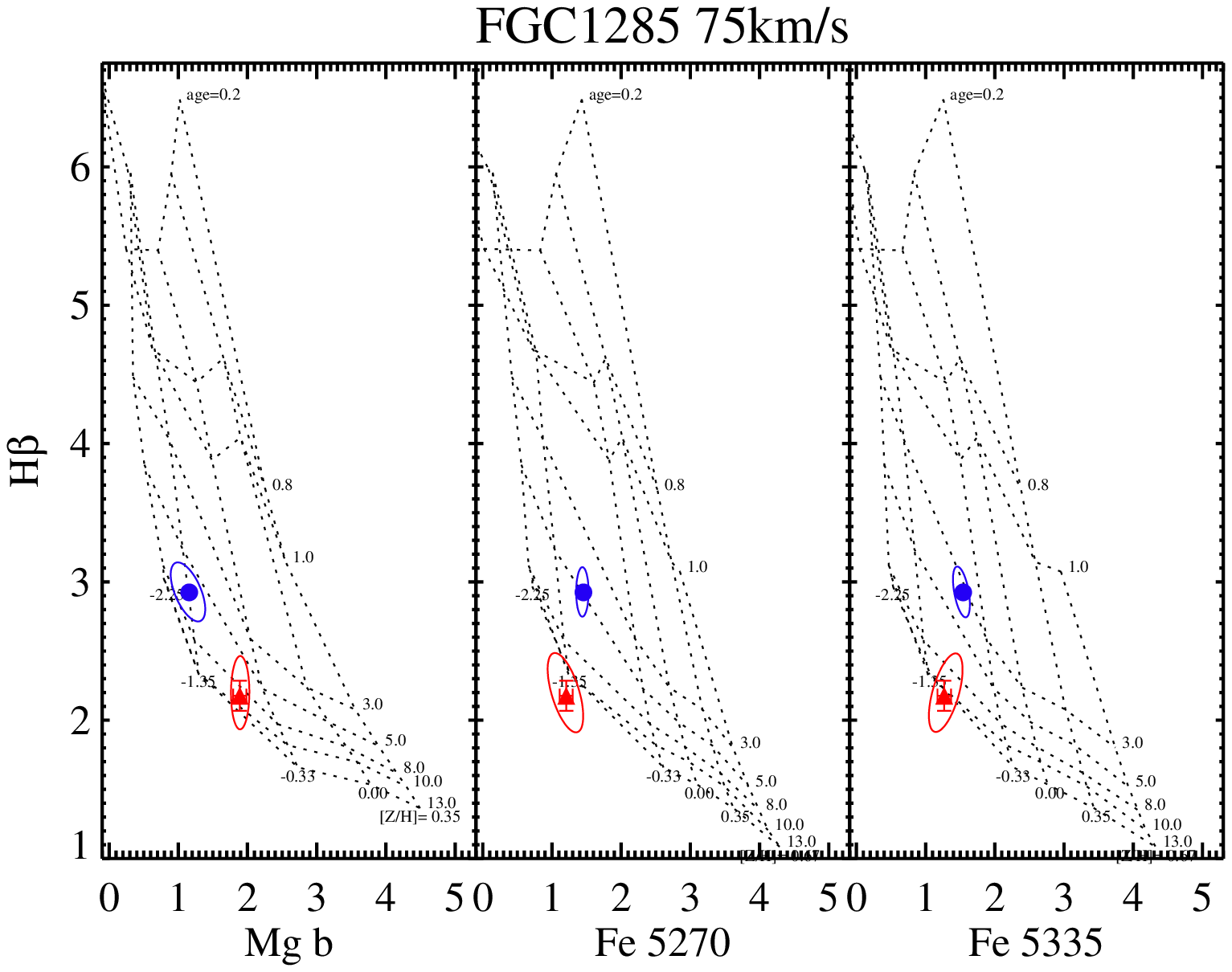}\\
\plotone{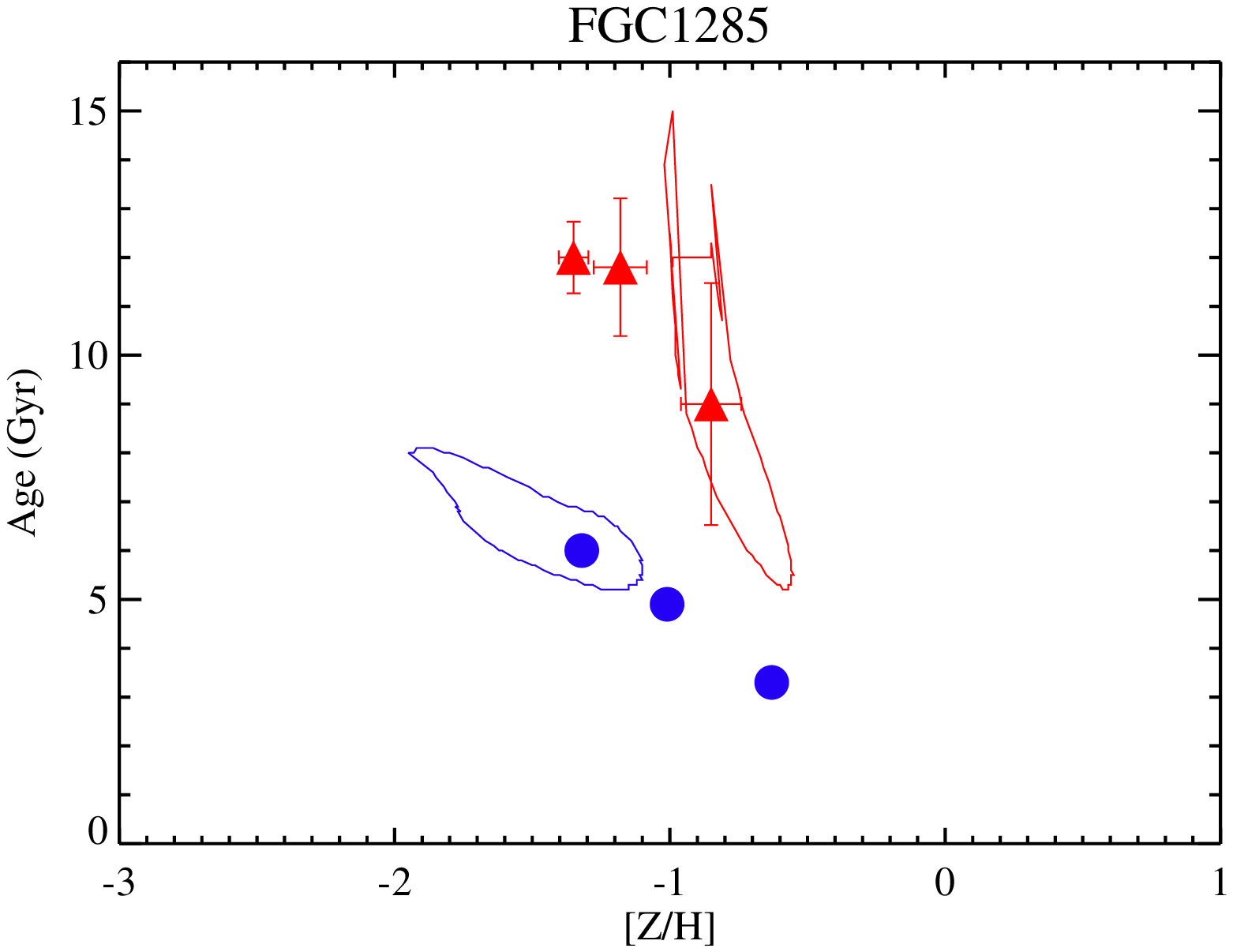}\\
}
\centerline{Fig. 5. --- Continued.}
{
\plotone{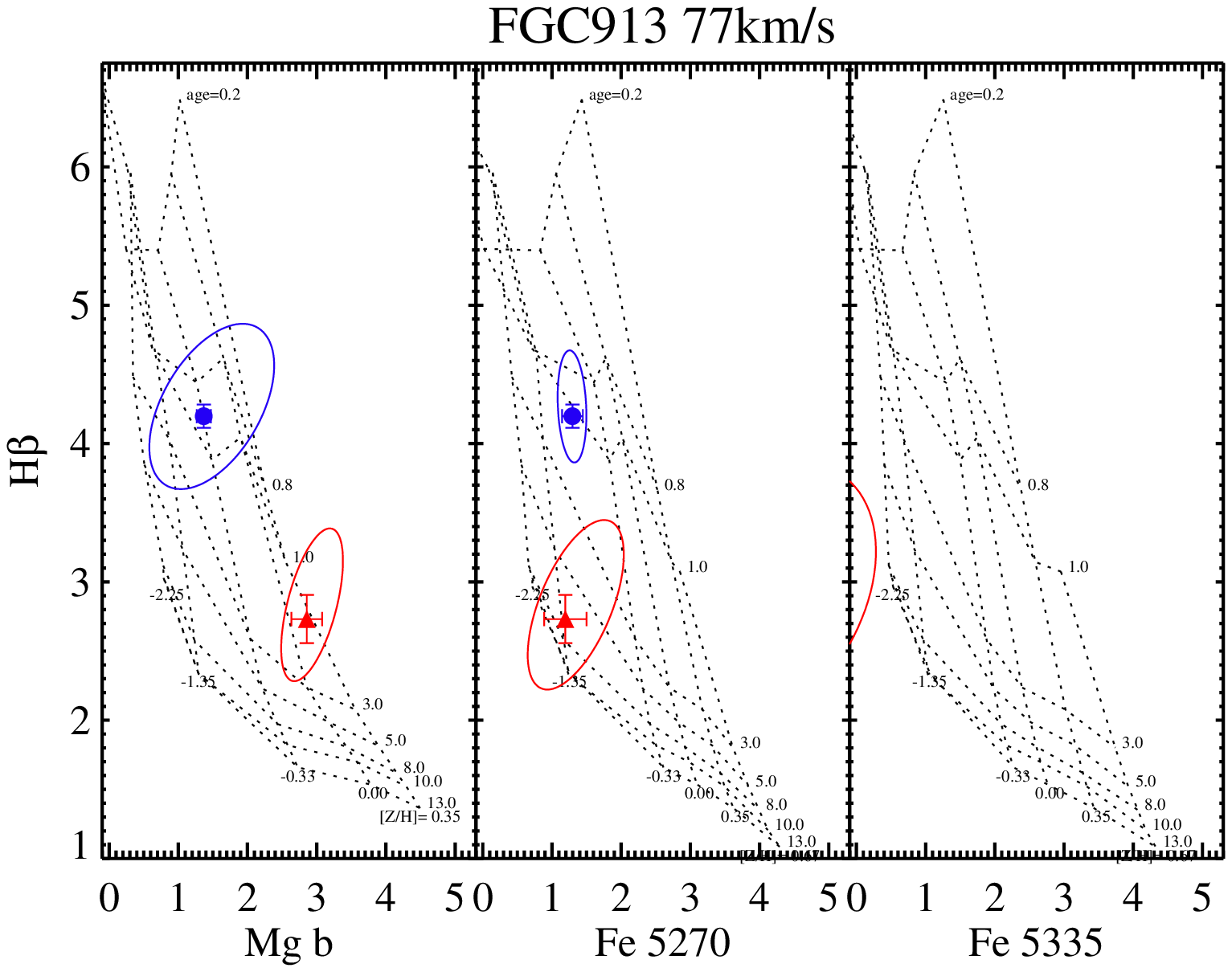}\\
\plotone{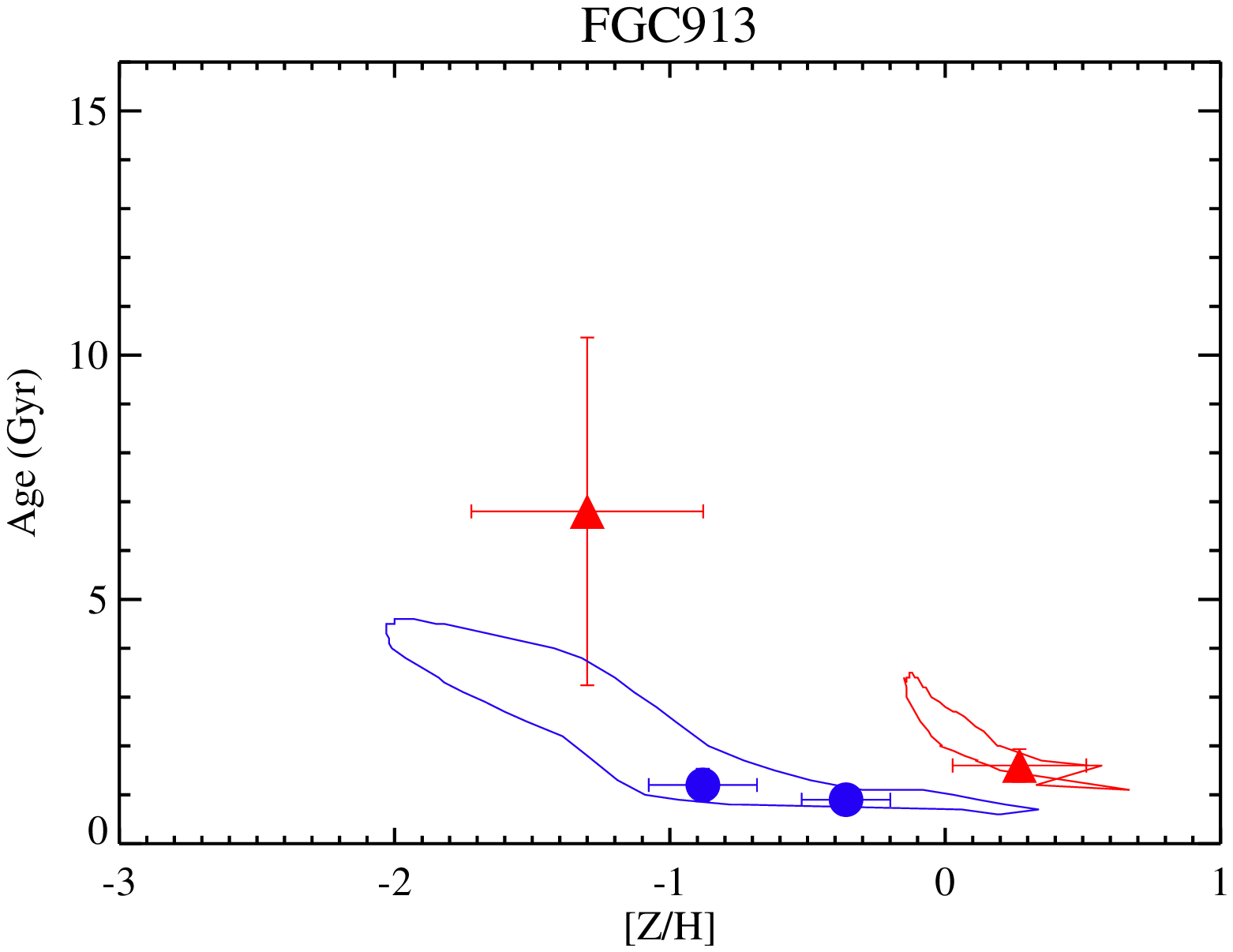}\\
}
\centerline{Fig. 5. --- Continued.}
{
\plotone{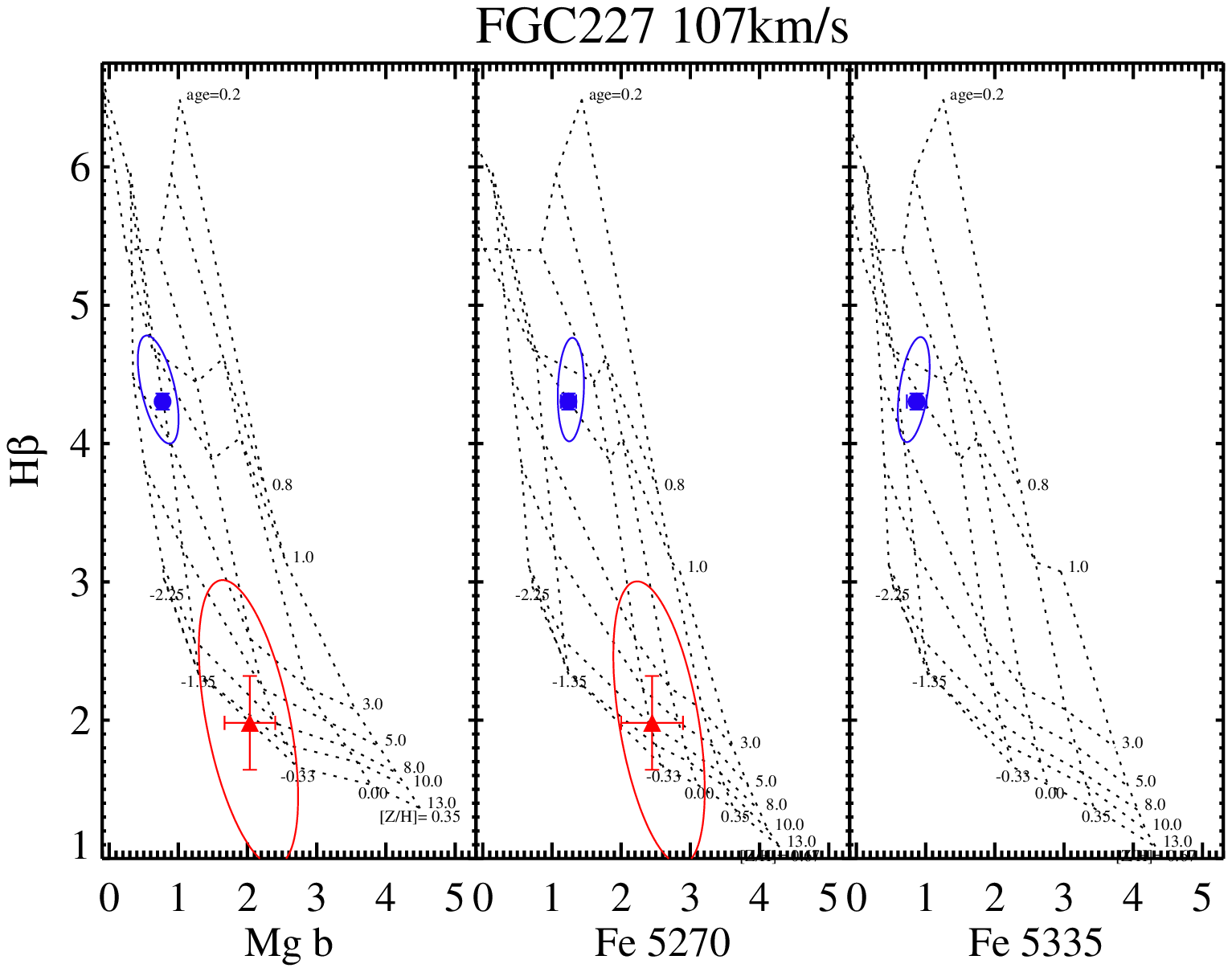}\\
\plotone{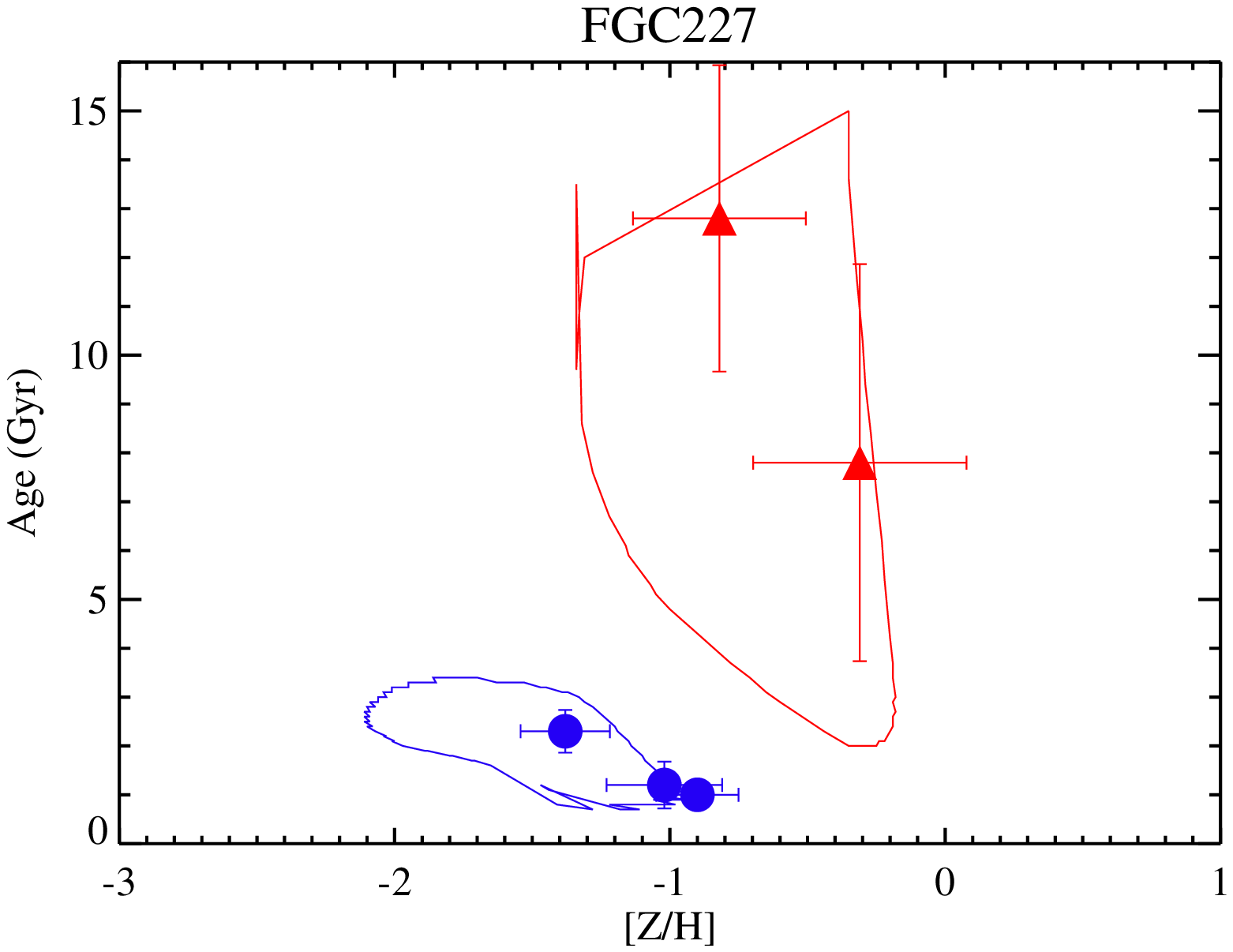}\\
}
\centerline{Fig. 5. --- Continued.}
{
\plotone{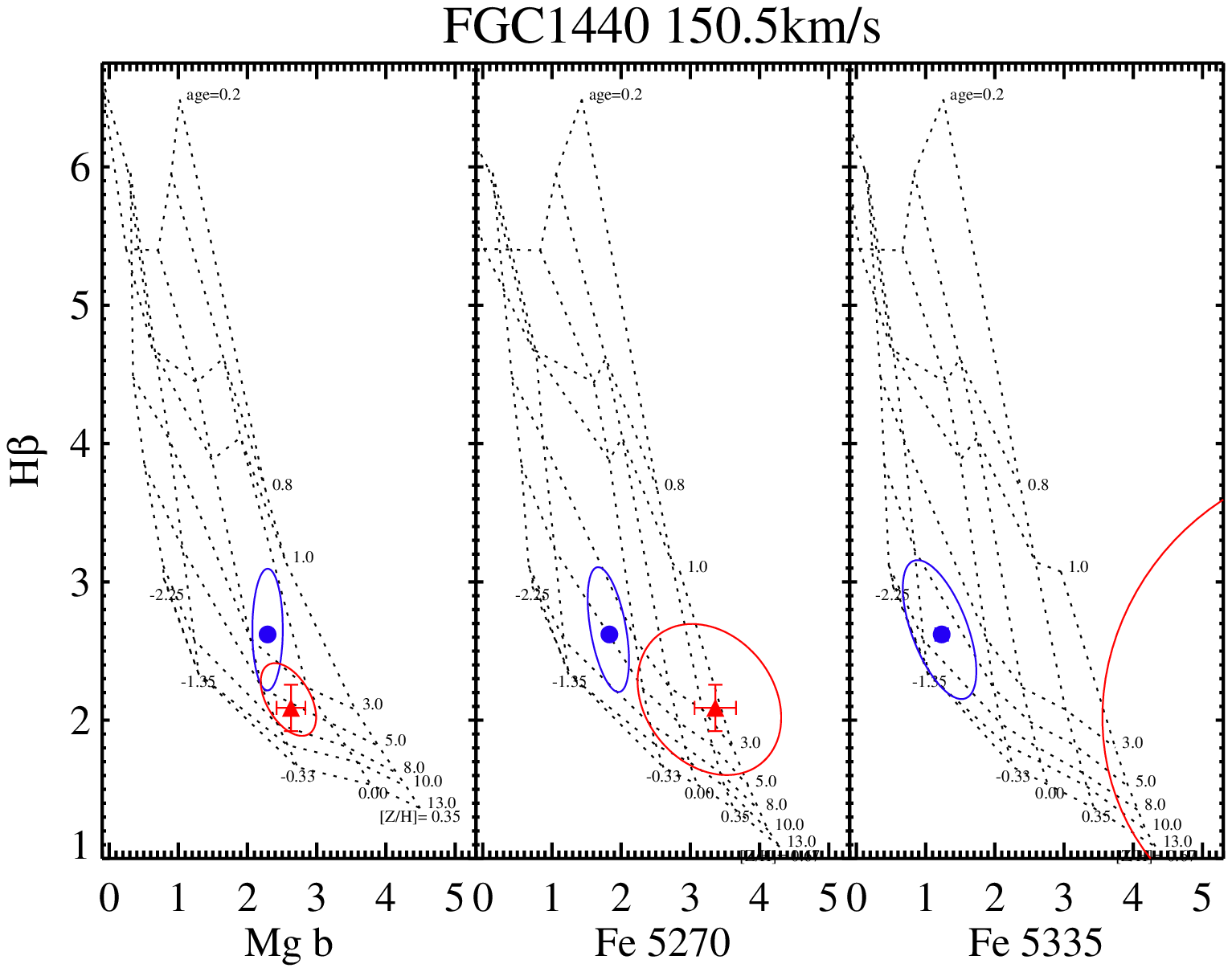}\\
\plotone{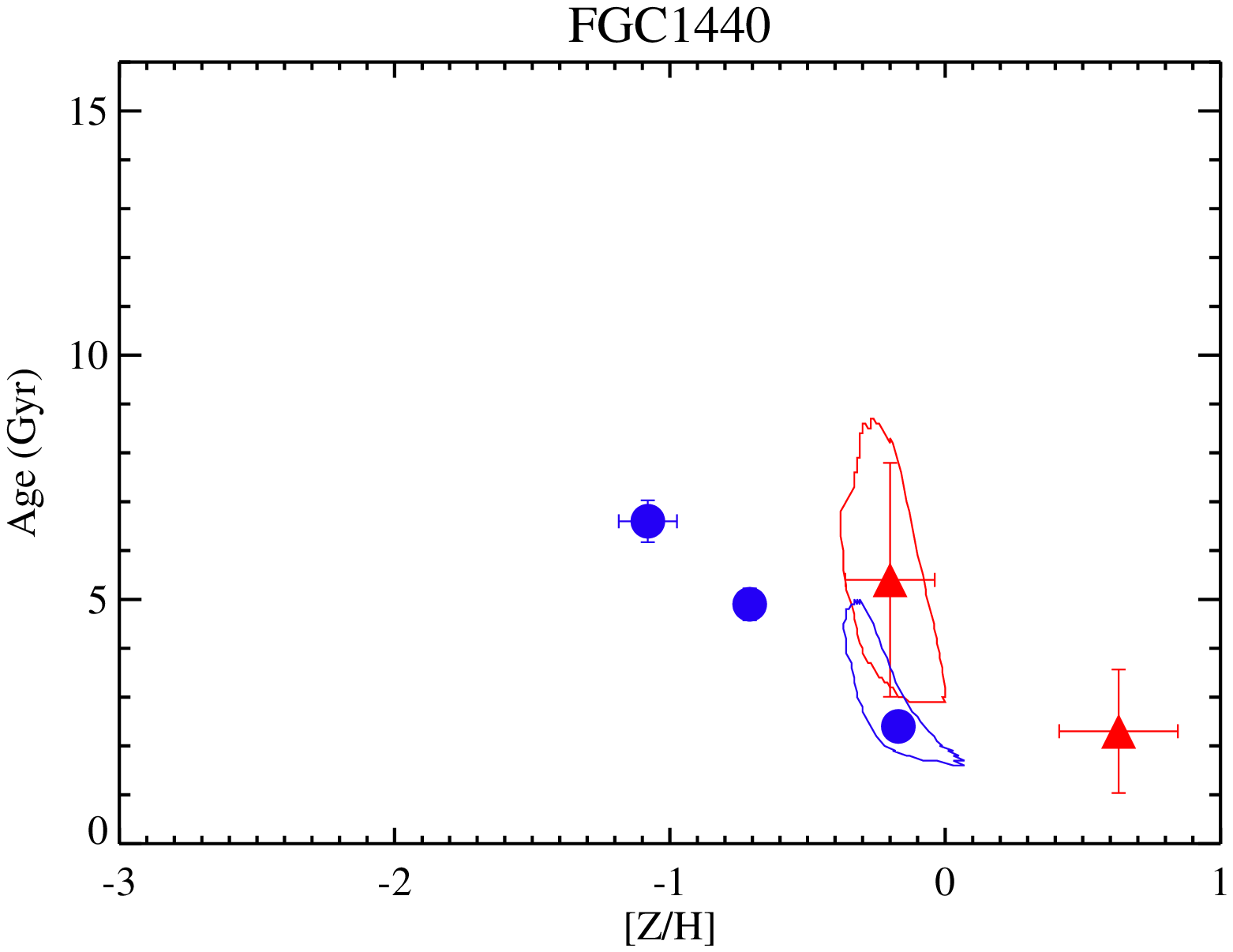}\\
}
\centerline{Fig. 5. --- Continued.}

\subsection{Radial Gradients}\label{sec_radgrad}

\begin{figure}
\plotone{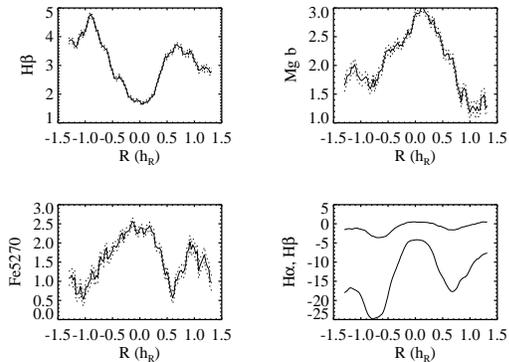}
\caption{ Radial gradients of the Lick equivalent-widths measured in FGC 1440.  Top left shows the \hb\ index after correcting for emission line fill-in, top right shows \mgb, lower left shows the Fe5270 index, and the lower right shows the uncorrected \hb\ EW as well as the \ha\ EW.  Dotted lines show the uncertainties calculated from the extracted spectra SNR. \label{rad_ind}}
\end{figure}

\begin{figure}
\plotone{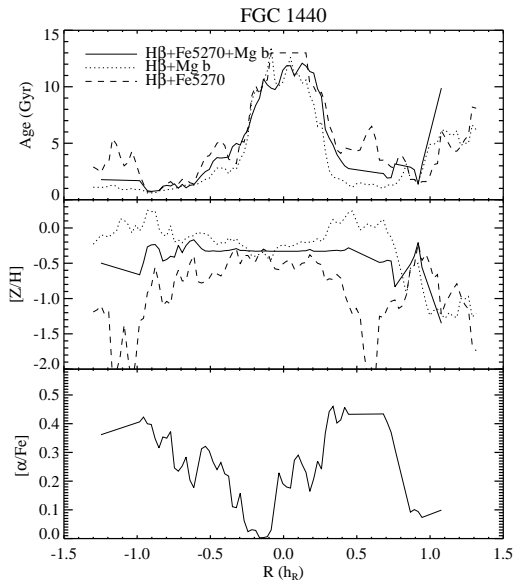}
\caption{ The interpolated age and metallicity measured along the midplane of FGC 1440.  The solid line shows the best fit when the \hb, \mgb, and Fe~5270 indices are simultaneously fit to the best matching \citet{Thomas03} model.  Dashed and dotted lines show the best fit if a solar composition is assumed.  There is a very strong age gradient present.  The \mgb\ index shows a fairly flat metallicity gradient, while the Fe~5270 index shows a radially decreasing metallicity.  The third panel shows the best fitting $\alpha$-enhancement when we simultaneously fit all three indices.  \label{rad_age}}
\end{figure}

The observations of the midplane of FGC 1440 were deep enough that we
can extract radial gradients of the Lick indices.  We extracted
spectra by binning 6.3\arcsec\ spatially in a sliding region across
the galaxy, and moving the observations onto the Lick system as
before.  This gives us a SNR of 200 in the central regions of the
galaxy and 50 at 1.3 $h_R$.  The radial variation of the Lick indices
along with the interpolated ages and metallicities are plotted in
Figures~\ref{rad_ind} and~\ref{rad_age}.  The \hb, \mgb, and Fe 5270
indices have radial gradients of 2.8, -1.5, and -1.3 \AA/$h_R$
respectively.  These gradients are fairly large for pure disk
systems. \citet{MacArthur06} finds similar gradients with similar
magnitudes in four late-type face-on systems.  \citet{Moorthy06} also
find strong radial gradients, but attribute them to the transition
between bulge and disk dominated regions of their galaxies.

The gradients in the Lick indices are consistent with a large age
gradient, with the central region of FGC~1440 having an SSP age of
$~\sim12$ Gyr dropping to 2 Gyr after one scale length.  The
metallicity gradients, however, are slightly ambiguous.  The
metallicity measured from the \mgb\ index shows a mostly flat radial
gradient, with a large drop on only one side of the galaxy.  On the
other hand the Fe~5270 index shows a decreasing metallicity on both
sides of the galaxy.  Fitting a line to the radial data, we find that
the \mgb\ index has a metallicity gradient of $-0.40 \pm 0.1$
dex/$h_R$.  If we restrict the fit to the region inside one
scale length, the \mgb\ metallicity gradient is consistent with
zero. The Fe~5270 index reveals a much steeper gradient of $-0.70 \pm
0.1$ dex/$h_R$.  This seems to imply that there is a radially changing
level of $\alpha$-enhancement throughout the galaxy, with the central
regions being close to solar composition and the outer regions
becoming more $\alpha$-enhanced, thereby inflating the metallicity
measured from \mgb.  In Figure~\ref{rad_age}, we plot the best fitting
$\alpha$-element enhancement and it does appear that the outer regions
are $\alpha$-enhanced compared to the central region which is best fit
with a nearly solar chemical composition.  These $\alpha$-element
measurements should be regarded with caution, as we have fit a model
with three free parameters (age, metallicity, and $\alpha$ composition)
using only three measurements (\hb, \mgb, and Fe 5270).  There is also
the possibility that we should adopt a radially varying emission line
correction.  This seems likely, as FGC 1440 hosts a dust lane which
becomes less prominent with radius.

While we cannot draw broad conclusions based on a single galaxy, it is
clear we detect stronger radial gradients than have been found in
other disk systems.  This is even more surprising given that edge-on
projection effects should act to smooth any radial population
gradients we observe.  We discuss these results further in \S\ref{rcg}.


\subsection{Possible Uncertainties}
\subsubsection{Emission Line Correction}\label{elcorr}

Both the midplane and offplane have prominent emission lines.  For the
midplane spectra, we applied \hb\ corrections that had an average of 4.9 \AA,
while the offplane corrections had an average of 2.2 \AA.  If we had
not take reddening or continuum shape into account and naively used
just the case B recombination correction for both the midplane and
offplane, we would have derive slightly younger ages for the midplane.
This simple correction would place many of the observations near the
upper edges of the model grids, suggesting we were under correcting.
If instead we assume an even higher amount of dust extinction for the
midplane, several of the thin and thick disks would then have similar
SSP ages.  However, it would take fairly extreme levels of dust
extinction to drive all of the thin disks to have similar ages as the
thick disks.  The blue broad-band colors for these galaxies are
inconsistent with such high levels of dust.

Propagating an uncertainty of $\pm10\%$ in the \hb\ emission
correction results in a median $\mp$0.2 dex shift in metallicity and
$\pm$1.9 Gyr in age for the midplane and $\mp$0.1 dex and $\pm$2.8 Gyr
for the offplane.

\subsubsection{Cross Contamination}

While we have placed our longslits at regions that should be dominated
by the thin or thick components, we expect some thick disk stars to be
present in the midplane and vice-versa.  Using the photometric fits in
\citet{Yoachim06}, we find that our midplane observations typically
contain $\sim$20\% thick disk flux, while the offplane observations
have a flux contribution of $\sim$25\% from the thin disk.  As can be
seen in Figure~\ref{slit_place}, we did not have large gaps between
our slit positions.  Observations made in poor seeing conditions might
therefore experience extra cross-contamination as light from the
midplane could be scattered into the offplane slit position. This should
not be a major problem as we avoided making offplane observations
during the worst seeing conditions, but could slightly increase the
amount of expected cross-contamination.

\citet{Serra07} study how Lick indices and their derived SSP ages and
metallicities are affected when there are multiple stellar populations
present.  They find that the derived ages are very sensitive to the
youngest stars present, while the metallicity measures are
predominantly influenced by any older population.  Contamination could
therefore explain the similar metallicities we measure for the thin
and thick disk stars.  However, resolved stellar population studies
have found small vertical metallicity gradients in low mass galaxies
as well \citep{Seth05b}.  If enough thick disk stars contaminate the
midplane, our SSP derived metallicities will be slightly biased
towards those of the older population, even in the young midplane.  Of
course, it is also possible that these low mass galaxies have simply
not undergone substantial star formation episodes and thus have not
chemically enriched the thin disks above the level of the thick disk.
Overall, the effects of cross-contamination would lead us to
underestimate the true metallicity differences between the thin and
thick disks, but to overestimate the flux-weighted age differences.

\subsubsection{Complex Stellar Populations}

While we are measuring SSP-equivalent ages and metallicities, it is
fairly obvious that the midplanes of disk galaxies have undergone
multiple epochs of star formation and are not well described by a
single age and metallicity.  Our ages and metallicities are thus best
interpreted as flux-weighted averages across the extracted radial
region, particularly for our midplane spectra where we have evidence
for very steep radial age gradients in some systems.  Because we are
forced to bin our spectra over a large spatial region to reach
adequate SNR, we include flux from the younger outer regions of the
galaxy.  Our midplane ages should probably be interpreted as minimums,
as the central region is undoubtedly older than the flux averaged
measure we report.

\subsubsection{$\alpha$-element Enhancement}

Many spectroscopic observations of elliptical galaxies and spiral
bulges have found stellar populations that have systematic 
differences between the metallicity calculated from the \mgb\ index
compared to the Fe indices.  This systematic shift is usually
interpreted as being caused by a stellar population that is
significantly enhanced with $\alpha$-elements compared to the spectra
that were used in building the Lick model grids.  Such an enhancement
is expected for stellar populations that form rapidly ($<1$ Gyr) and
that are primarily enriched by Type II supernovae \citep{Matteucci94}.
Enhancements in $\alpha$-elements are often seen in elliptical
galaxies \citep{Worthey92,Fisher96,Thomas03}, as well as in local MW
thick disk stars \citep{Bensby05}.  

The galaxies in our sample are all
fairly low mass and therefore also low metallicity.  In the low
metallicity regime, the signature of $\alpha$-element enhancement
becomes weaker in the Lick indices.  Unlike massive elliptical
galaxies where the metallicity indicators can show systematic offsets
of $\sim$0.5 dex for an $\alpha$-enhanced population, our galaxies are
all sub-solar metallicity and thus would show little bias even if they
are $\alpha$-enhanced.  If we used model SSP grids with
[$\alpha$/Fe]=0.3, our derived metallicities would change by only
$\sim0.1$ dex.

We are hesitant to use our data to fit the $\alpha$-element
enhancement level.  If we forge ahead and do so, we find considerable
spread between the metallicities returned, but neither the thin or
thick disk have systematically larger metallicities returned from the
\mgb\ index, as we would expect if the stars were $\alpha$-enhanced.
However, given the small expected offset and lower SNR than available
for elliptical galaxy spectra, we do not consider this a significant
result, and include it here only for completeness.

\section{Discussion}

\subsection{Are Low Mass Thick Disks Old?}

Measuring accurate ages for thick disk stars has been done in
relatively few systems.  In the MW, stars that are kinematically
identified as thick disk stars typically have ages greater than 8 Gyr
\citep{Fuhrmann98,Bensby04b}.  HST studies of resolved stellar
populations in nearby galaxies show the offplane regions are dominated
by old stars.  \citet{Seth05b} find that in 8 galaxies the offplane
RGB stars have ages in the range of 2-6 Gyr.  Similarly,
\citet{Mould05} uses the ratio of RGB and AGB stars to find ages of
thick disks in a sample of 4 galaxies to be older than 3 Gyr.  Our
measured thick disk SSP ages fall between 3.8 and 10.9 Gyr with a
median age of 7.1 Gyr, consistent with these other studies that show
thick disks to be dominated by ancient stars.

\subsection{Are Low Mass Thick Disks Metal Poor?}

Many of our thick disks appear to be more metal rich than the embedded
thin disks.  This counter-intuitive result is probably due to the flux
weighted nature of our measurement.  The young, metal-poor outer
regions dilute the true central ages and metallicities of the thin
disks.

Thick disk metallicities have only been measured for a handful of
systems.  MW thick disk stars typically have metallicities in the
range [Fe/H]$\sim-0.7$ to -0.2, with the highest metallicity thick
disk stars possibly reaching solar values \citep{Bensby06}.  One
difficulty with comparing to the MW thick disk is that the observed
MW properties are of thick disk stars near the solar radius, while we
have only been able to measure thick disk properties near the central
regions of the galaxies.  Fortunately there are signs that the MW
thick disk has relatively small age and metallicity gradients
\citep{Bensby05}.

Constraints of thick disk metallicities have also been derived from
HST studies imaging resolved stars.  \citet{Seth05b} used ACS images
of 6 nearby edge-on disks to constrain vertical gradients in the
stellar populations using the color and distribution of AGB and RGB
stars.  The older RGB stars have a systematically larger scale height
compared to the younger AGB and main sequence stars.  They find little
to no metallicity gradients in the thick disk stars in their systems,
with the metallicities of the thick disks peaked around
[Fe/H]$\sim-1$.  This is slightly more metal poor than the
measurements we have for our thick disks.  However, the offset is
unlikely to be significant, given that the \citet{Seth05b} study is
able to study a cleaner sample of thick disk stars by reaching higher
vertical heights which reducing the contamination of thin disk stars.
Like the data presented here, \citet{Seth05b} only studies lower mass
systems, limiting the amplitude of any possible metallicity gradient
due to the low metallicity of the midplane.  Using similar HST
observations, \citet{Mould05} finds that thick disk stars in 4 edge-on
galaxies have [Fe/H] between -1.0 and -0.78, again very similar to the
metallicities we find.

\begin{figure}
\plotone{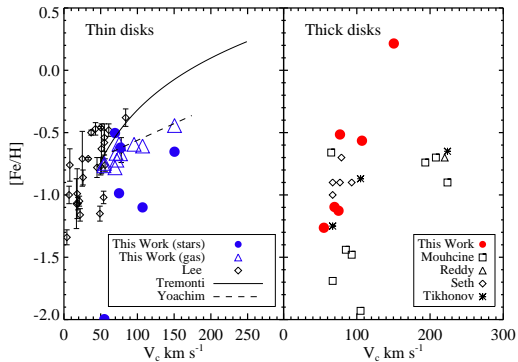}
\caption{Comparison of our thin and thick disk metallicities to
similar studies.  The panel on the left shows our thin disk
metallicities along with the low mass galaxies in \citet{Lee06b} and
the fits to larger samples in \citet{Tremonti04} and
\citet{Yoachim07b}. On the right, we compare our thick disk
metallicities to the systems measured in \citet{Seth05b},
\citet{Mouhcine05}, \citet{Reddy06}, and
\citet{Tikhonov05}. \label{comp} }
\end{figure}

In Figure~\ref{comp}, we compare our thin disk metallicities to the
low mass sample in \citet{Lee06b} and the large SDSS sample of
\citet{Tremonti04}.  We also compare our thick disk values to thick
disk and halo samples presented in
\citet{Seth05b,Mouhcine05,Tikhonov05}, and \citet{Reddy06}.  In cases
where the authors presented values of log(O/H)+12, we converted to
[Fe/H] assuming log(O/H)$_\odot$+12=8.69 and [Fe/O]=0.  We also plot
the midplane nebular abundances for our galaxies calculated from the
S2N2 calibrator \citep{Viironen07}.  With the exception of a few
outliers, our measured metallicities are consistent with metallicities
measured in similar systems.

Recently, \citet{Ivezic08} have questioned if the Milky Way thick disk
is a unique component or simply an extension of the thin disk that has
non-Gaussian metallicity and velocity distributions \citep{Norris87}.
They cite a lack of correlation between velocity and metallicity in
large SDSS samples as a major problem for traditional disk
decompositions and rule out a ``traditional" two-disk model at the
8$\sigma$ level.  Unfortunately, the model \citet{Ivezic08} rule out
is not applicable to their observations.  In particular, they model
1,142 stars observed in the region $1.0 < z/kpc < 1.2$.
\citet{Ivezic08} correctly assume the observations will contain a
similar number of thin and thick disk stars, however they do not use a
realistic thin disk component.  The thin disk stars are modeled as
having an asymmetric drift of 9 \kms and [Fe/H]=-0.50 with a spread of
0.04 dex (their Figure~16).  These parameters would be appropriate for
modeling nearby thin disk stars, but they are observing stars 3-4 thin
disk scale heights above the Galactic plane.  At this large height,
only the kinematically hottest thin disk stars will be present in the
sample, and one should expect a much larger velocity lag than observed
in local cooler thin disk stars.  \citet{Holmberg07} show that the
hottest local disk stars are the oldest and that the age-metallicity
relation (AMR) for local stars has large intrinsic dispersion at large
ages ($\sigma\sim0.2$ dex).  This large metallicity dispersion is
probably a result of radial migration of the older high velocity
dispersion stars \citep{Haywood08}.

If the offplane thin disk stars are assumed to be similar to older
local thin disk stars (i.e., kinematically hot with a broad
metallicity distribution and larger asymmetric drift), their properties
should well match SDSS observations of regions at large scale height.
Specifically, there should be little correlation between kinematics
and metallicity despite the mixture of thin and thick disk stars.
This revised model is consistent with the observations of thick disks
presented here and elsewhere \citep{Seth05b} that find stars at large
scale heights are a significantly older population than those found
near the midplane.


\subsection{Are Thick Disks $\alpha$-enhanced?}

The level of $\alpha$-enhancement can be a major clue to the formation
process of a stellar population.  If $\alpha$-enhanced, it is a sign
that a stellar population has been enriched mostly over a short time
period by Type II SNe, whereas stars with solar composition formed
over an extended period and have been enriched by both Type II and
Type Ia SNe.  Numerous papers have found that MW thick disk stars are
enhanced in $\alpha$-elements compared to thin disk stars at similar
total metallicities \citep[e.g.,][]{Bensby03, Bensby04a, Bensby05,
Feltzing03, Reddy06, Taut01,mashonkina03, Prochaska00, Fuhrmann98,
Fuhrmann04}.  Unfortunately, we do not have the SNR to definitively
say if our thick disks are $\alpha$-enhanced, due to the overall low
metallicities of our target galaxies.

\subsection{Are Thick Disks ``Normal'' Stellar Populations?}

Using a sample of $>$1000 SDSS images of edge-on disk galaxies,
\citet{Zibetti04} examine the faint halo that appears when the images
are stacked.  This extended halo has anomalous colors, requiring stars
that are either metal rich or have a bottom heavy IMF
\citep{Zibetti04,Zackrisson06}.  We find no such anomalies with the
thick disks we observe spectroscopically, as most of them fall on
stellar synthesis model grids using standard IMFs and metallicities.
The few galaxies where we measure Lick indices that are inconsistent
with the SSP models can easily be explained as spurious measurements
caused by low SNR, and do not require exotic stellar populations.

\subsection{Radial Color Gradients in the Thin Disks}\label{rcg}

\citet{Bell00} observed broadband colors for a large sample of
galaxies and found that the radial gradients were predominantly caused
by age gradients in the stellar populations.  By averaging their
sample together, They found a metallicity gradient, but broadband
sensitivity to dust makes this a measurement in individual galaxies.
Their work has been expanded on by \citet{MacArthur04}, who found both
metallicity and age gradients are stronger in the inner regions of
galaxies, and that galaxies with strong age gradients had smaller
metallicity gradients.  The Lick indices ability to lift the
age-metallicity degeneracy, and relative insensitivity to dust makes
it much easier to quantify how much of the radial color gradients in
disk galaxies are due to age versus metallicity changes.

In one of the only study that has explicitly targeted Lick indices in the
disk-dominated regions of galaxies, \citet{MacArthur06} detect
age gradients in only 2 of their 8 galaxies, and find strong negative
metallicity gradients in 4.  The age gradients in \citet{MacArthur06}
are also rather small (-0.5 and -1.3 Gyr/$h_R$).  Our measurements of
FGC 1440 show a much steeper age gradient, with the SSP age dropping
by $\sim9$ Gyr over one scale length.

\citet{Ganda07} use the SAURON integral field unit spectrograph to
measure Lick index strengths across the face of 18 late-type disk
galaxies.  They find in general \hb\ increasing with radius and the
metal sensitive lines decreasing with radius.  The SAURON observations
show a very wide range of galaxy-to-galaxy radial behavior, the most
extreme are consistent with the strong gradients we find in FGC 1440.

Our finding that the thin disk of FGC 1440 might have stellar
populations of near solar composition and be $\alpha$-enhanced at
larger radii is puzzling.  Looking at the transition between bulge and
disk dominated regions, \citet{Moorthy06} finds that the central
bulges are either solar-composition or $\alpha$-enhanced, with little
to no $\alpha$-enhancement in the disks.  With the presence of old
thick disks in all of the galaxies, we would expect the galaxies to
have experienced plenty of chemical evolution and enrichment from SNe
Ia.  Instead, the outer regions of FGC 1440 are $\alpha$-enhanced,
suggesting that the central region of the galaxy has undergone
extended chemical enrichment, while the outer regions have not, despite
being surrounded by old thick disk stars.

This could be a sign that the thick disk stars in FGC 1440 have been
recently accreted, and thus have not contributed to the chemical
enrichment the galaxy.  Another possibility is that the central region
of the galaxy is the only place where the gravitational potential is
deep enough to retain SN ejecta, and the outer disk has historically
suffered from SN blow-out and failed to retain metal enriched gas.
Another possibility is that the luminosity-weighted metallicity in the
outer disk is dominated by enrichment from the latest burst of
star-formation which has $\alpha$-enhanced the region.

\section{Conclusions}

We have spectroscopically confirmed that the thick disks observed in
edge-on late type galaxies are old, metal-poor stellar populations,
analogous to the thick disk stars seen in the MW and nearby edge-on
systems.  This is the first time ages and metallicities of thick disks
have been measured in unresolved stellar populations.  Because all of
our targets are fairly low mass, we are unable to detect any
significant differences between thin disk and thick disk
metallicities.  After correcting for emission line contamination, the
thin disks in our sample are found to be quite young, with strong
radial age gradients.

We fail to detect any significant trend for thick disk stars to be
enhanced in $\alpha$-elements compared to their thin disks which is a
defining characteristic of the MW thick disk.  Our failure to observe
$\alpha$-enhancements is most likely a result of our sample being
dominated by low-mass and therefore low metallicity galaxies, for
which solar and $\alpha$-enhanced models are similar.

For one galaxy in our sample we have measured the radial gradients of
the Lick indices in the thin disk and find the large gradients that
are dominated by changes in the average stellar age with a small
contribution from a changing average metallicity.

\acknowledgments

We thank Connie Rockosi and the UW machine shop for helping in the
design and manufacture of our slit.  We also thank the APO observing
specialists for their help executing the observations.  We thank
Suzanne Hawley for reading an early version of this paper and making
helpful comments.  JJD and PY were partially supported through NSF
grant CAREER AST-0238683 and the Alfred P.\ Sloan Foundation.  Based
on observations obtained with the Apache Point Observatory 3.5-meter
telescope, which is owned and operated by the Astrophysical Research
Consortium.


\end{document}